\documentclass[aps,prd,reprint,a4paper,showpacs,nofootinbib,superscriptaddress]{revtex4-2}

\usepackage{bbm}
\usepackage{amsmath}
\usepackage{graphicx}
\usepackage{float}
\usepackage{amssymb}
\usepackage{subfigure}
\usepackage{amssymb}
\usepackage{hyperref}
\usepackage{epstopdf}

\usepackage[utf8]{inputenc}
\hypersetup{
    colorlinks=true,
    linkcolor=red,
    citecolor=blue,
}
\usepackage{color}
\usepackage[T1]{fontenc}
\usepackage{txfonts}
\usepackage{mathrsfs}
\usepackage{latexsym}
\usepackage{epsfig}
\usepackage{epstopdf}
\usepackage{epstopdf}
\usepackage{graphicx}
\usepackage{amssymb}
\usepackage{amsmath}
\usepackage{dcolumn}
\usepackage{bm}
\usepackage{color}
\usepackage{comment}
\usepackage{xcolor}
\usepackage{dsfont}

\begin{document}
\title{\bf Magnetized particle motion and accretion process with shock cone morphology around a decoupled hairy black holes}

\author{G. Mustafa}
\email{gmustafa3828@gmail.com }
\affiliation{Department of Physics, Zhejiang Normal University,
Jinhua 321004, China}

\author{Faisal Javed}
\email{faisaljaved.math@gmail.com}\affiliation{Department of
Physics, Zhejiang Normal University, Jinhua 321004, People's
Republic of China}

\author{S.K. Maurya}
\email{sunil@unizwa.edu.om}\affiliation{Department of Mathematical and Physical Sciences, College of Arts and Sciences, University of Nizwa, Nizwa 616, Sultanate of Oman}

\author{A. Ditta}
\email{adsmeerkhan@gmail.com}
\affiliation{Research Center of Astrophysics and Cosmology, Khazar University, Baku, AZ1096, 41 Mehseti Street, Azerbaijan}

\author{Orhan~Donmez}
\email{orhan.donmez@aum.edu.kw}
\affiliation{College of Engineering and Technology, American University of the Middle East, Egaila 54200, Kuwait}

\author{Tayyab Naseer}
\email{tayyabnaseer48@yahoo.com}\affiliation{Department of
Mathematics and Statistics, The University of Lahore,
1-KM Defence Road Lahore-54000, Pakistan}

\author{Abdelmalek Bouzenada}
\email{abdelmalekbouzenada@gmail.com}
\affiliation{Laboratory of Theoretical and Applied Physics, Echahid Cheikh Larbi Tebessi University 12001, Algeria}

\author{Farruh Atamurotov}
\email{atamurotov@yahoo.com}
\affiliation{Urgench State University, Kh. Alimdjan str. 14, Urgench 220100, Uzbekistan}

\begin{abstract}
Relativistic accretion onto compact objects such as black holes and neutron stars is one of the most efficient known mechanisms for converting gravitational potential energy into radiation. In the case of rapidly spinning black holes, up to $40\%$ of the rest-mass energy of accreting matter can be released, far exceeding the efficiency of nuclear fusion. In this work, we investigate magnetized particle motion and relativistic accretion processes around a decoupled hairy black hole via extended geometric deformation. The developed geometry involves two hairy parameters that preserve the horizon structure with the additional feature of the fulfillment of weak energy conditions outside the event horizon. We provide the foundation with necessary formalism for magnetized particle motion around a decoupled black hole. The effective potential and innermost stable circular orbits are then derived, which demonstrate a significant reduction of the radius of the latter quantity under the hairy parameters for the magnetized particle. Afterwards, we obtain exact analytical expressions for radial velocity profiles, mass accretion rates, and a few others which reveal improved energy efficiency and emissivity as compared to the standard black hole. Furthermore, the decoupling parameter shows strong influence on oscillations, accretion presenting fantastic agreement between analytical predictions and numerical simulations, and thus offering noticeable observational signatures for future gravitational wave and X-ray astronomy.

\end{abstract}

\maketitle

\date{\today}

\section{Introduction}

According to predictions of the theory of general relativity (GR) \cite{mmB}, black holes (BHs) are fascinating cosmic entities that serve as formidable producers of gravitational fields and have the potential to have tremendous spin and magnetic strength. These environments provide a unique setting for the investigation of the interactions that occur between gravity and matter in astrophysical settings. The existence of BHs has been confirmed by recent discoveries in the field of observation. These discoveries include the detection of gravitational waves resulting from the merger of two BHs by the LIGO and Virgo collaboration, the first image of the shadow of the BH in $M87^{\ast}$, and the most recent image of the BH in the center of the Milky Way, Sagittarius $A^{\ast}$, which was captured by the Event Horizon Telescope using very long baseline interferometry \cite{1,2,3,4}. Furthermore, electromagnetic emissions that are produced by accretion disks that surround BHs are an essential component in the detection of these objects \cite{5,6,7}. These substantial breakthroughs not only bolster our understanding of general relativity and the characteristics of accretion disks surrounding supermassive BHs, but they also provide insights into the effects of strong gravity in the vicinity of the event horizon, which has the potential to shed light on modified theories of gravity.

An accretion disk around a BH is a revolving disk comprised of gas and dust that arises from the gravitational pull of the BH, drawing in neighboring matter. Comprised mostly of highly energized gas, this disk releases intense radiation as it approaches the event horizon of the BH. An accretion disk is essential for understanding the dynamics and energy release processes associated with BHs. The continuous advancement in accretion theory, propelled by accurate and analytical answers, has enabled a deeper understanding of many astrophysical situations, assisting in the comprehension of the processes being studied \cite{34df}. Moreover, these analytical solutions are crucial reference points for validating numerical models, which makes them essential instruments in this domain \cite{35df}. According to Newtonian gravity, the Bondi model describes the continuous descent of a spherically symmetric fluid into a BH \cite{36df}. Michel used a Schwarzschild BH as a reference point to expand this concept to the field of relativity \cite{38df}. Bondi and Hoyle created analytical answers for cases involving wind acceleration in Newtonian theory \cite{37df}, and Tejeda and Aguayortiz found analytical solutions for a Schwarzschild BH \cite{38df}. Our understanding of spherical and wind accretion processes has been improved by many studies that used analytical and computational methods \cite{40f,41f,42f,43f,44f,45f}. References \cite{t1f}-\cite{t14f} include full details of investigations into different BH spacetimes.

Magnetically charged BH \cite{47f}. To analyze the accretion process for charged BHs of higher dimensions, Sharif and Iftikhar \cite{48f} established the generalized equation for the relativistic Bernoulli equation and mass, as well as the conservation of energy flux. In addition, the study of the motion of the particles around rotating BH is presented in \cite{49f}. In the presence of massless and massive scalar fields, the dynamical configuration of different BH geometries is presented in \cite{50af}-\cite{50bh}. Ditta and Abbas \cite{50f} investigated circular orbits around the regular phantom BH. In addition, they presented a detailed study of the accumulation of matter for various BHs in \cite{51f}. They analyzed the process of steady-state accretion of ideal fluids onto a regular Hayward BH and found that the rate at which matter accumulates onto a typical Hayward BH differs from that of a Schwarzschild BH. The speed of sound is inversely correlated with the increase in the length parameter and is also directly proportional to the radius \cite{52f}. A study of the geodesic structure and accretion behavior around a BH, as described by non-linear electrodynamics, is reported in the reference \cite{rr52f}. The study considers both strong and weak field approximations. These estimations demonstrate the emergence of a disk-shaped configuration as the BH undergoes geodesic motion and accretion.

The most recent advances in the field of BH studies have sparked a new understanding of the effects of quantum gravity, the morphology of dark matter, and the observational traits of strong gravity. The effects of quantum and spacetime curvature on the structure and associated phenomenology of BHs have been modeled. The non-minimally coupled scalar field and Hawking radiation of regular BHs have been analyzed \cite{v1}. The scenario wherein the primordial regular BHs, arising from quantum covariant effective gravity or non-time-reversal-symmetric spacetimes, are proposed to be dark matter candidates \cite{v2,v3}. Further, researchers describe the possible effects of dark energy on the cosmological BHs and the gravitational waves resulting from them \cite{v4}. The Event Horizon Telescope has provided evidence against certain modified gravity theories like mimetic gravity \cite{v5}. The development of theories on quasinormal radiation and BH spacetime has allowed the use of BH shadows as a probe of fundamental physics \cite{v6}. These works include detailed calculations of strong field gravity near Sgr A* \cite{v7}, constraints on the BH superradiance phenomenon and associated shadow ultralight particles \cite{v8}, \cite{v9}, and detailed shadow studies containing scalar hair \cite{v10}. These works show the increasing collaboration and harmony of quantum gravity inspired BHs and observational precision across the astrophysical spectrum.

Furthermore, research has been conducted on modified theories of gravity, including the non-commutative theory \cite{54f}, quantum gravity corrections \cite{55f}, and acceleration processes on scalar-tensor-vector gravity \cite{53f}. Furthermore, investigations have been conducted to investigate the phenomenon of heteroclinic and cyclic accretion to BHs with potential $f(R)$ and $f(T)$ \cite{56f,57f}. Various studies have been conducted using dynamical systems and BH thermodynamics to investigate the phenomenon of accretion flows into stationary BHs, as documented in citations \cite{58f}-\cite{63f}. In addition, researchers have investigated other alternative BH geometries, such as those described in \cite{p1f}-\cite{p11f}. Through the use of observational data from a wide range of compact objects, Liu and his \cite{rr53f,rr54f} colleagues have successfully computed and tested quasi-periodic oscillations.

Accurate solutions to the equations of motion are essential for understanding the intricacies of gravitational physics. These solutions form a foundation for investigating the dynamics of massive systems, enabling scientists to examine strong field conditions and analyze the nonlinear interactions that shape our universe. Various methods have been documented in the literature for deriving these solutions, such as applying specific configurations of the curvature tensor, utilizing symmetry groups, and employing numerical methods. The significance of these solutions is evident in their role in validating numerical simulations and enhancing our understanding of gravitational interactions.
Additionally, analytical solutions serve as a foundation for perturbative approximations or ansatz techniques (selecting specific metric potentials), which facilitate a more profound comprehension of the underlying physical phenomena. The search for exact solutions is an essential aspect of the ongoing effort to uncover various enigmas related to gravity and the cosmos, with the existing literature providing a diverse array of methods and perspectives to support this exploration \cite{1a00}–\cite{1g00}.

The recently proposed gravitational decoupling method through minimal geometric deformation by Ovalle \cite{2900} has proven to be an effective approach to address field equations in alternative gravity theories. By enabling this decoupling, the method systematically generates viable anisotropic solutions, which are essential for accurately modeling compact astrophysical objects, such as neutron stars and BHs. The existing literature underscores the significance of this technique in examining how modified theories, such as Einstein-Weyl gravity, affect the dynamics of celestial systems. Ovalle and Linares \cite{3000} expanded their research by analyzing an isotropic spherically symmetric geometry within the braneworld framework. They successfully derived an exact solution consistent with the Tolman-IV ansatz, showcasing the adaptability of their methodology. This foundational work paved the way for further investigations, including the extension of Casadio et al.'s approach to derive the Schwarzschild geometry in a braneworld context \cite{3100}. Furthermore, Ovalle and his team \cite{3300} explored the transition from a spherical isotropic fluid interior to a viable anisotropic configuration through this method. Sharif and Sadiq \cite{3400} advanced this research by applying their method to charged configurations, presenting two new anisotropic forms of Krori-Barua components and performing a graphical analysis. They established a range of physically viable anisotropic systems using the Duragpal–Fuloria geometry as a foundational isotropic structure \cite{3600,36z00}. Graterol \cite{37fa00} also contributed by introducing innovative solutions that involved deformations in the radial Buchdahl coefficient. This technique has also been used to derive physically acceptable anisotropic solutions for Heintzmann and Tolman VII at various parameter values \cite{36a00,37a00}. 
This scheme has yielded several impressive results; however, it remains an incomplete strategy. This limitation arises because the above-mentioned method does not accommodate the structure of a BH featuring a clearly defined horizon. Consequently, Ovalle \cite{35d01} implemented transformations in both radial and time metric functions to derive anisotropic solutions, a process he referred to as extended geometric deformation (EGD). Several studies have recently discussed BH solutions using these well-known techniques \cite{t7f}-\cite{t11f}. Numerous other significant studies employing these strategies have been conducted, showing the flexibility and extensive applicability of this recently introduced systematic scheme \cite{3500}-\cite{11d00}. 

In this paper, new nonlinear electrodynamic BHs are defined, and new accretion results inspired by general relativity (GR) are derived. Here is the outline of the paper: Section (\ref{sec2}) introduces the spacetime of non-linear electrodynamic BHs. The velocity of the test particles is extensively calculated in Section (\ref{ssp}). Next, we will explore oscillations and circular motion. Section IV examines the parameters of a perfect fluid, including its critical flow speed, mass accretion, and its temporal history. In Section (\ref{se5}), we take a look at the physical consequences of strong and weak fields of these results. In Section (\ref{se51}), the paper concludes, with an emphasis on how parameters $\beta$ and $Q$ greatly affect both strong and weak field approximations. The spacetime signature $(-,+,+,+)$ and the geometric units $G=c=1$ are used throughout the investigation.

\section{New decoupled BH solution} \label{sec2}

\begin{figure*}
\centering \epsfig{file=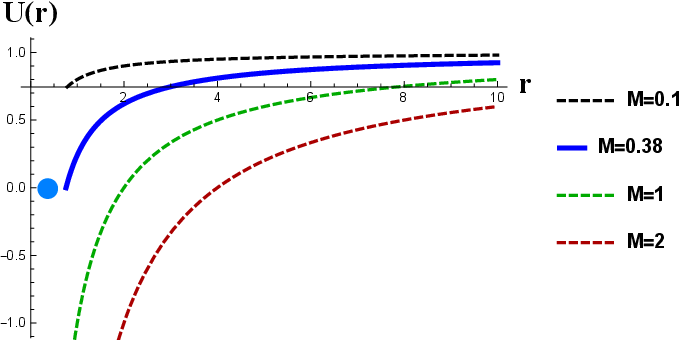, width=.42\linewidth,
height=2in}~~~~~~~~\epsfig{file=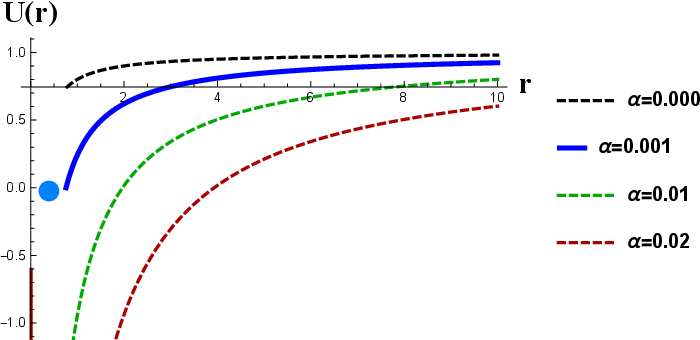, width=.42\linewidth,
height=2in}\\
\vspace{0.5cm}
\centering \epsfig{file=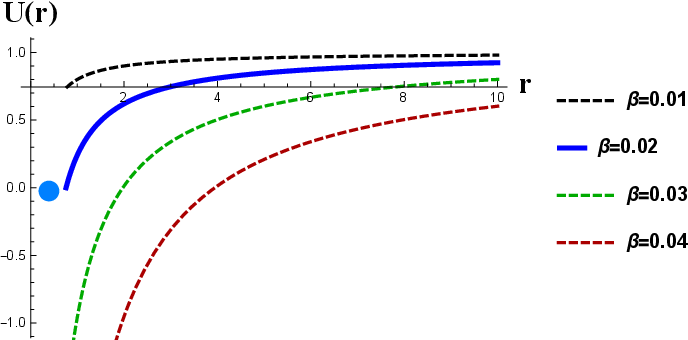, width=.42\linewidth,
height=2in}
\caption{\label{F1} Theory concerning the variation of the horizon of a decoupled BH with respect to parameters $M$, $\alpha$, and $\beta$. The changes in mass $M$ and the characteristics of the BH $\alpha$ change the color of the event horizon to blue. For any mass $M$ and Hairy parameters $\alpha$ and $\beta$, all solution curves in the display lack inner and outer horizons.}
\end{figure*}

The present section discusses the newly formulated BH solution by Avalos and his colleagues \cite{zz1}, applying the Einstein field equations under the concept of gravitational decoupling in the following application:
\begin{eqnarray}\label{1a}
    G_{ij} \equiv R_{ij}-\frac{1}{2} R g_{ij}=\kappa^{2} \tilde{T}_{ij},
\end{eqnarray}
where $\kappa^{2}$ equals $\frac{8 \pi G}{c^4}$ and $\tilde{T}_{ij}$ is the total EMT, derived from two different sources of matter, defined as:
\begin{eqnarray}\label{2a}
\tilde{T}_{ij}=T_{ij}+\hat{T}_{ij}. 
\end{eqnarray}
In Eq. (\ref{2a}) above, $T_{ij}$ represents a standard EMT, whereas $\hat{T}_{ij}$ represents the additional material distribution linked with the earlier fluid makeup under gravity. It is important to mention that the standard EMT follows the Bianchi identity, leading to the covariant conservation principle expressed as:
\begin{eqnarray}\label{3a}
    \nabla_i \tilde{T}^{ij}=0.
\end{eqnarray}

At present, through the study of a stationary system affected by spherically symmetrical attributes, which is defined as:
\begin{eqnarray}\label{4a}
    d s^2=e^{\mathcal{F}_0(r)} d t^2-e^{\mathcal{H}_0(r)} d r^2-r^2 d \Omega^2,
\end{eqnarray}
where, both gravitational functions, $\mathcal{F}_0=\mathcal{F}_0(r)$ and $\mathcal{H}_0=\mathcal{H}_0(r)$, depend only on the radial coordinate $r$ and $d\Omega^2=d\theta^2+\sin^2\theta d\phi^2$. The calculations for the Einstein field equations are determined within the framework of the decoupling scheme given by:
\begin{eqnarray}
&&\hspace{-0.5cm}\frac{1}{\kappa^{2}}\left[\frac{1}{r^2}-e^{-\mathcal{H}_0}\bigg(\frac{1}{r^2}-\frac{\mathcal{H}_0 '}{r} \bigg)\right] =  T_0^0+\hat{T}_0^0 ,\label{T0}\\
&&\hspace{-0.5cm}\frac{1}{\kappa^{2}}\left[\frac{1}{r^2}-e^{-\mathcal{H}_0}\bigg(\frac{1}{r^2}+\frac{\mathcal{F}_0 '}{r} \bigg)\right]=  T_1^1+\hat{T}_1^1 ,\label{T1}\\
&&\hspace{-0.5cm}-\frac{e^{-\mathcal{H}_0}}{4\kappa^{2}}\left[2\mathcal{F}_0 '' +\mathcal{F}_0'^2-\mathcal{H}_0' \mathcal{F}_0' +\frac{2\mathcal{F}_0'-2\mathcal{H}_0'}{r} \right]=T_2^2+\hat{T}_2^2.\label{T2}
\end{eqnarray}
We have established the concept of an efficient matter sector by establishing connections between the matter variables tied to the initial and additional fluid matter distribution. They are expressed as:
\begin{eqnarray}
\rho&=&T_0^0+\hat{T}_0^0,\label{5a}\\
p_r&=&-T_1^1-\hat{T}_1^1,\label{6a}\\
p_t&=&-T_2^2-\hat{T}_2^2.\label{7a}
\end{eqnarray}
Now, we apply EGD to the initial metric contained in the source $\hat{T}_{ij}$ as follows:
\begin{eqnarray}
\mathcal{A}_0 \rightarrow \mathcal{F}_0&=&\mathcal{A}_0+g, \label{10a}\\
e^{-\mathcal{F}_0}  \rightarrow e^{-\mathcal{H}_0}&=&e^{-\mathcal{S}_0}+f\label{11a},
\end{eqnarray}
$f$ and $g$ represent the components of geometric deformations. By examining Eqs. (\ref{10a}) and (\ref{11a}) within Eqs. (\ref{T0})-(\ref{T2}), the Einstein field equations can be described, leading to two different situations. The initial formulation of field equations is described as:
\begin{eqnarray}
\frac{1}{\kappa^{2}}\left[\frac{1}{r^2}-e^{-\mathcal{S}_0}\bigg(\frac{1}{r^2}-\frac{\mathcal{S}_0 '}{r} \bigg)\right]&=&T_0^0,\label{T0seed}\\
\frac{1}{\kappa^{2}}\left[\frac{1}{r^2}-e^{-\mathcal{S}_0}\bigg(\frac{1}{r^2}+\frac{\mathcal{A}_0 '}{r} \bigg)\right]&=&T_1^1,\label{T1seed}\\
-\frac{e^{-\mathcal{S}_0}}{4\kappa^{2}}\bigg(2\mathcal{A}_0 '' +\mathcal{A}_0'^2-\mathcal{S}_0' \mathcal{A}_0' +2\frac{\mathcal{A}_0'-\mathcal{S}_0'}{r} \bigg)&=&T_2^2.~~~~\label{T2seed}
\end{eqnarray}
The system \eqref{T0seed}-\eqref{T2seed} can be solved by the following components:
\begin{eqnarray}\label{8a}
d s^2=e^{\mathcal{A}_0(r)} d t^2-e^{\mathcal{S}_0(r)} d r^2-r^2 d \Omega^2,
\end{eqnarray}
with its mass function represented by the following formula:
\begin{eqnarray}\label{9a}
e^{-\mathcal{S}_0(r)} \equiv 1-\frac{\kappa^{2}}{r} \int_0^r x^2 T_0^0(x) d x=1-\frac{2 m(r)}{r}. 
\end{eqnarray}
The mass function of the seed geometry under consideration is defined as $m=m(r)$ in the relation provided above. On the other hand, new field equations for the $\Theta_{ij}$ sector are introduced with identical transformations as: 
\begin{eqnarray}
 \frac{1}{\kappa^{2}}\left[- \frac{f}{r^2}-\frac{f'}{r}\right]=\hat{T}_0^0,\label{T0source}\\
 \frac{1}{\kappa^{2}}\left[- X_1 - f \bigg( \frac{1}{r^2}+\frac{\mathcal{F}_0'}{r} \bigg)\right]=\hat{T}_1^1,\label{T1source}\\
 \frac{1}{\kappa^{2}}\left[- X_2 - \frac{f}{4} \bigg( 2\mathcal{F}_0''+
\mathcal{F}_0'^2+2\frac{\mathcal{F}_0'}{r}\bigg)\right]-
 \frac{f'}{4\kappa^{2}}\bigg( \mathcal{F}_0' +\frac{2}{r} \bigg)=\hat{T}_2^2,\label{T2source}
\end{eqnarray}
the values of $X_1$ and $X_2$ are defined as:
\begin{eqnarray}
X_1&=&\frac{e^{-\mathcal{S}_0} g'}{r}, \\
X_2&=&\frac{e^{-\mathcal{S}_0}}{4} \bigg( 2g''+ g'^2 +2\frac{g'}{r}+2\mathcal{A}_0 ' g' -\mathcal{S}_0' g' \bigg).
\end{eqnarray}
The first part of the mentioned field equations corresponds to the standard energy-momentum tensor, labeled as $T_{ij}$. The second choice, guided by a condition based on physical reality, makes it easier for us to talk about the $\Theta$-sector. The appropriate decoupling of the system is a result of the energy interaction between the two matter suggestions, $T_{ij}$ and $\hat{T}_{ij}$ \cite{zz1}. After verifying the transfer of energy, one can continue evolving a hairy BH by making modifications to the Schwarzschild metric, which serves as the starting point. The Schwarzschild metric is stated in the following equation:
\begin{eqnarray}\label{12a}
1-\frac{2 M}{r}=e^{\mathcal{A}_0}=e^{-\mathcal{S}_0}.
\end{eqnarray}
When deformation functions $\{f, g\}$ are carefully chosen, the event horizon resulting from the deformed solution can be defined as:
\begin{eqnarray}\label{13a}
e^{\mathcal{F}_0\left(r_H\right)}=e^{-\mathcal{S}_0\left(r_H\right)}=0,
\end{eqnarray}
$r_H$ represents the correlation for the radius of the horizon. In Schwarzschild coordinates, there is a relationship that exists:

\begin{eqnarray}\label{14a}
    e^\mathcal{F}_0=e^{-\mathcal{S}_0},
\end{eqnarray}
with the constraint given by:
\begin{eqnarray}\label{15a}
    \tilde{p}_r=-\tilde{\rho}.
\end{eqnarray}
The geometric distortion function $f$ can be defined in relation to additional variables by using Eq. (\ref{14a}) in the following relation:
\begin{eqnarray}\label{16a}
f=\left(1-\frac{2 M}{r}\right)\left(e^g-1\right),
\end{eqnarray}
which leads to:
\begin{eqnarray}\label{17a}
d s^2= \left(1-\frac{2 M}{r}\right) H(r) d t^2-\left(1-\frac{2 M}{r}\right)^{-1} H^{-1}(r) d r^2-r^2 d \Omega^2.
\end{eqnarray}
 We will add another constraint to find the auxiliary function $H(r)$. In \cite{zz1}, it was found that one can observe other scenarios, such as strong energy conditions, both inside and outside the BH's boundary. It is important to mention that BHs must also meet a less strict condition called weak energy requirements to be considered physically significant \cite{zz2}. This work focuses primarily on the issues associated with inadequate energy conditions. Mathematically, they are expressed as follows:
\begin{eqnarray}\
&&0\leq\tilde{\rho}, \nonumber \\
&&0\leq\tilde{\rho}+\tilde{p}_r,\label{WEC}\\
&&0\leq\tilde{\rho}+\tilde{p}_t. \nonumber
\end{eqnarray}
In particular, Eq. (\ref{15a}) in conjunction with the preceding statements yields the following:
\begin{eqnarray}
0\leq\hat{T}_0^0 , \label{WEC1} \\
\hat{T}_2^2\leq\hat{T}_0^0 .\label{WEC2}
\end{eqnarray}
By employing Eqs. (\ref{T0source}) and (\ref{T2source}), the previous inequalities can be restated as:
\begin{eqnarray}
0\leq 1-H(r)-(r-2M)H'(r) , \label{ineq1} \\
0\leq 2-2H(r)+4MH'(r)+r(r-2M)H''(r) .\label{ineq2}
\end{eqnarray}
Currently, there are certain functions $H(r)$ that satisfy Eq. (\ref{ineq1}) if:
\begin{eqnarray}
G(r)=1-H(r)-(r-2M)H'(r),\label{eqa3}
\end{eqnarray}
for $G(r)>0$. The general solution for the equation mentioned above is as follows:
\begin{eqnarray}
\frac{r-c_1}{r-2M} - \frac{1}{r-2M} \int G(x)dx=H(r).
\end{eqnarray}
In the equation mentioned earlier, $c_1$ represents the constant with length dimensions. Furthermore, there is a further requirement regarding $G(r)$ due to constraint (\ref{ineq2}), defined as:
\begin{eqnarray}\label{G}
0\leq 2 G(r)-r G'(r).
\end{eqnarray}
We may assert that $G(r)$ is an arbitrary positive function that satisfies Eq. (\ref{G}). In the present analysis, we see it this way:
\begin{eqnarray}
G(r)=\alpha \frac{M}{r^2} \ln \bigg( \frac{r}{\beta} \bigg),
\end{eqnarray}
the replacement in Eq. (\ref{eqa3}) yields:
\begin{eqnarray}
1-H(r)-(r-2M)H'(r)=\alpha \frac{M}{r^2} \ln \bigg( \frac{r}{\beta} \bigg) \geq 0, \label{ODE} 
\end{eqnarray}
where $\alpha$ and $\beta$ are constant with lengths as their dimensions, given the following condition:
\begin{eqnarray} \label{alfacondition}
0\leq \alpha. 
\end{eqnarray}
The exact solution of Eq. (\ref{ODE}) is:
\begin{eqnarray}\label{ODEsolution}
H(r)=\frac{r-c_1}{r-2M} + \frac{\alpha M}{r(r-2M)} \bigg(1+\ln \bigg( \frac{r}{\beta}\bigg) \bigg).
\end{eqnarray}
Setting $c_1$ equal to $2M$ yields the initial Schwarzschild solution in the limit where $\alpha$ approaches zero. The inequality (\ref{ineq2}) is satisfied in this case given by:
\begin{eqnarray}\label{betac}
\ln\bigg( \frac{r}{\beta}\bigg)\geq \frac{1}{4}.
\end{eqnarray}
By replacing Eq. (\ref{ODEsolution}) with (\ref{17a}), we obtain the resulting solution, which is already reported in \cite{mainarticle} in the form of:
\begin{eqnarray}\label{1}
U(r)= e^\mathcal{F}_0=e^{-\mathcal{S}_0}=1-\frac{2M}{r}+\frac{\alpha M}{r^2}+\frac{\alpha M}{r^2} \ln \bigg( \frac{r}{\beta} \bigg).
\end{eqnarray}

In the above equation, the newly obtained solution's lapse function affected by gravitational decoupling is represented as $U(r)$. We aim to bring about numerous concerns to light. Gravitational decoupling expands on the Schwarzschild BH solution by incorporating a supplementary source $\hat{T}_{ij}$, while upholding the fundamental theory. The demonstration of the onset of this phenomenon is displayed in various parametric values $\alpha$ and $\beta$. The Schwarzschild BH solution is achieved as $\alpha$ approaches zero. Moreover, more reasons for this can be seen in \cite{mainarticle}. The impact of the decoupling properties of the BH is very noticeable and indivisible within the field limit $U(r)$. The lapse function is visually represented in Fig. (\ref{F1}). The blue curve at $M=0.38$ in the top left Fig. represents the event horizon, with fixed values of $\alpha=0.001$ and $\beta=0.01$. The event horizon is shown in the top right graph as a blue curve at $\alpha=0.001$, with constant values for $M=1$ and $\beta=0.01$. The event horizon is shown in the lower graphic as a blue curve at $\beta=0.02$, with fixed values of $M=1$ and $\beta=0.001$. There are no inner or outer horizons in the horizon structure of decoupled BHs, regardless of the mass $M$ and the parameters $\alpha$ and $\beta$.

\begin{figure*}
\centering \epsfig{file=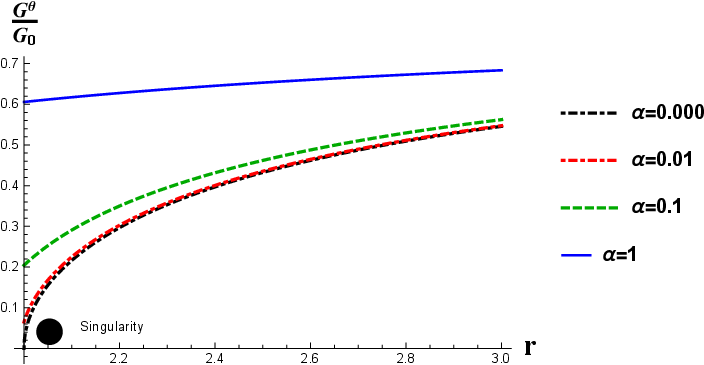, width=.4\linewidth,
height=2.0in}~~~~~~~~\epsfig{file=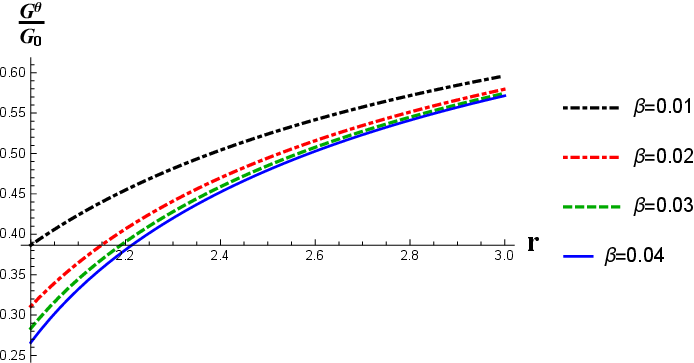, width=.4\linewidth,
height=1.7in}\caption{\label{F3} Changes in magnetic field caused by variations in independent parameters $\alpha$ and $\beta$.In both plots, the magnetic field strength grew as $\alpha$ increased and $\beta$ decreased away from the singularity. Although these graphs have an important role in the magnetic field model, they only show minor changes in the decoupled parameters.}
\end{figure*}

{\color{black}\section{Magnetized particle motion around decoupled BH} \label{ssp}

We examine a decoupled geometric BH surrounded by a uniformly distributed magnetic field. Considering that the magnetic field is static, uniform and radially symmetric with a strength $ G_0 > 0 $ around the BH, we assume that the field is weak enough not to alter the spacetime geometry beyond the BH. In case  of a magnetic field  stronger, the spacetime geometry would need to be modified to account for its influence. Since the spacetime metric in (\ref{1}) accommodates both space-like and time-like Killing vectors, we can apply the Wald method \cite{round21} to determine the components of the electromagnetic field's 4-vector potential:
\begin{eqnarray}\label{4}
A_\phi=\frac{G_0r^2\sin^2\theta}{2}.
\end{eqnarray}
The non-zero components of the electromagnetic tensor can be determined using the formula $F_{\mu\nu} = A_{\nu, \mu} - A_{\mu, \nu}$ as:
\begin{eqnarray}\label{5}
F_{r\phi}&=& r G_0\sin^2\theta,\\
\label{6}
F_{\theta\phi}&=& r G_0\sin\theta \cos\theta.
\end{eqnarray}
For a magnetic field, the $4$-velocity $\mu_\mu$ can be determined as:
\begin{eqnarray}\label{7}
B^\mathcal{A}_0&=&1-\frac{1}{2}\mu_\mu F_{\eta\sigma}\delta^{\zeta\eta\sigma\mu},
\end{eqnarray}
where the Pseudo-tensorial form $\delta^{\eta\zeta\sigma\mu}$ is associated with the Levi-Civita symbol $\epsilon_{\eta\mathcal{A}_0\sigma\gamma}$ as:
\begin{eqnarray}\label{8}
\epsilon_{\eta\zeta\sigma\gamma}&=&\frac{\delta^{\eta\zeta\sigma\mu}}{\sqrt{-g}}, \quad \epsilon^{\delta\zeta\sigma\gamma}=\sqrt{-g}\delta_{\eta\zeta\sigma\mu},
\end{eqnarray}
where $g=r^4\sin^2\theta$. The magnetic field's non-zero components are calculated as
\begin{eqnarray}\label{9}
G^{\hat{r}} &=&\frac{G_0}{\sec\theta}, \quad G^{{\hat{\theta}}}=G_0\sin\theta\sqrt{U(r)}.
\end{eqnarray}
The behavior of the magnetic field, where  $\hat{G}^\theta$ as a function of the decoupling parameters $\alpha$ and $\beta$ is illustrated in Fig. (\ref{F3}). It is observed that the magnetic field increases as one moves away from the singularity when the value of $\alpha$ increases. For the case of large values of $\beta$. Next, we will analyze the motion of particles near a decoupled geometric BH under the influence of the magnetic field. The trajectories of these magnetized particles can be derived using the Hamilton-Jacobi equation \cite{gmr2}:
\begin{eqnarray}\label{9}
g^{\mu\nu}\frac{\partial N}{\partial x^\mu}\frac{\partial N}{\partial x^\nu}&=&-\Big(1-\frac{S^{\mu\nu}F_{\mu\nu}}{2m}\Big)^2m^2.
\end{eqnarray}
In this context, the variable $m$ represents the mass, and $N$ refers to the influence of magnetized particles within the BH's spacetime. The interaction between a magnetized particle and an external magnetic field is denoted by $S^{\mu\nu}F_{\mu\nu}$. The magnetic field of these particles is described by the polarization tensor $S^{\mu\nu}$ (for further details, see \cite{gmr2}):
\begin{eqnarray}\label{9a}
S^{\eta\zeta}&=&\delta^{\zeta\eta\epsilon\nu}\mu_{\nu}u_{\epsilon}, \quad S^{\zeta\eta}u_\eta=0.
\end{eqnarray}
In this expression, $\delta^{\zeta\eta\epsilon\nu}$ represents the Levi-Civita tensor, while $u^\mu$ denotes the 4-velocity, and $\mu^{\nu}$ is the four-vector of the magnetic dipole moment for magnetized particles. To define the electromagnetic field tensor $F_{\zeta\eta}$, it can be constructed using the magnetic field component $G^{\eta}$ and the electric field component $E^{\zeta}$:
\begin{eqnarray}\label{9b}
F_{\eta\zeta}&=&[\zeta^{E_\eta}]u-\delta^{\zeta\eta\epsilon\nu}u^\sigma G^\gamma.
\end{eqnarray}
By plugging Eq. (\ref{9a}) into (\ref{9b}), we get the interaction term $D* F$ as:
\begin{eqnarray}\label{9c}
D^{\zeta\eta}F_{\zeta\eta}&=&2\mu^\zeta\eta_\zeta=2\mu G_0\mathcal{L}[\lambda_{\hat{a}}].
\end{eqnarray}
In this context, $[\lambda_{\hat{a}}]$ represents a function of $r$, $\mu$ refers to the magnetic moment, and $[\lambda_{\hat{a}}]$ corresponds to the angular velocity \cite{gmr3}. Our study focuses on examining the orbital motion of magnetized particles around the dynamic phantom AdS BH.
When $\theta = \frac{\pi}{2}$ and $\dot{\theta} = 0$, the motion of magnetized particles in the equatorial plane can be described by:
\begin{eqnarray}\label{9d}
N=-Et+L\phi+N_r{r}+N_\theta{\theta}.
\end{eqnarray}
By substituting Eq. (\ref{9c}) into (\ref{9a}) and applying Eq. (\ref{9d}), the equation governing the radial motion of magnetized particles can be obtained:
\begin{eqnarray}\label{9e}
\dot{r^2}+V_{eff}(r;\alpha,\beta,L,\eta)=\zeta^2.
\end{eqnarray}
The magnetic coupling parameter, denoted as $\left(\eta=2\frac{G_{0}}{m\mu}\right)$, characterizes the interaction between a magnetized particle and an external magnetic field. In addition, $\left(L=\frac{L_1}{m}\right)$ and $\left(\zeta=\frac{E}{m}\right)$ represent the specific angular momentum and energy of the particle, respectively. The effective potential for magnetized particles is given by the following formula:
\begin{eqnarray}
V_{eff}=U(r)\Big(1-\mathcal{L}[\lambda_{\hat{a}}]\eta+\frac{L^2}{r^2}\Big).\label{13}
\end{eqnarray}
In this context, $\eta > 0$ indicates that the directions of the magnetic field and the magnetic dipole moment of the particle are aligned, while $\eta < 0$ signifies that they are opposite. When $\eta = 0$, it means that there is no external magnetic field or magnetic dipole moment present. A typical way to characterize the circular orbits of particles near a compact object is defined as follows:
\begin{eqnarray}\label{9f}
\dot{r}=0,~~~~~\frac{d}{dr}V_{eff}=0.
\end{eqnarray}
The magnetic coupling parameter $\eta$ for circular orbits can be determined using the initial condition from Equation (\ref{9f}) as follows:
\begin{eqnarray}\label{9g}
\eta(r;L,\zeta,\alpha,\beta)=\frac{1}{\mathcal{L}[\lambda_{\hat{a}}]}\Big(1-\frac{\zeta^2}{U(r)+\frac{L^2}{r^2}}\Big).
\end{eqnarray}
The second condition from Eq. (\ref{9f}) leads to the following:
\begin{eqnarray}\label{9h}
\frac{d}{dr}V_{eff}=U(r)\mathcal{L}[\lambda_{\hat{a}}]\frac{d}{dr}\eta.
\end{eqnarray}
We are interested in magnetic field components that are measured by an observer moving in frame with a particle.  This can be expressed as:
\begin{eqnarray}\label{9i}
G_{\hat{r}}=0=G_{\hat{\phi}},~~~~G_{\hat{\theta}}=G_0 U(r)e^\psi,
\end{eqnarray}
with the following condition:
\begin{eqnarray}\label{9j}
e^\psi=\frac{1}{\sqrt{U(r)-r^2\Omega^2}},
\end{eqnarray}
where $\Omega$ is the angular momentum, which is further defined as:
\begin{eqnarray}\label{9k}
\Omega=\frac{d\phi}{dt}=\frac{U(r)}{r^2}\frac{L}{\zeta}.
\end{eqnarray}
The interaction term can be determined from Eqs. (\ref{9a}), (\ref{9c}) and (\ref{9i}) as:
\begin{eqnarray}\label{9l}
D^{\zeta\eta}F_{\zeta\eta}&=&2\mu G_0U(r)e^\psi.
\end{eqnarray}
Now, by comparing Eq. (\ref{9c}) with (\ref{9l}), one can obtain the following relation:
\begin{eqnarray}\label{9m}
\mathcal{L}[\lambda_{\hat{a}}]&=& e^\psi U(r).
\end{eqnarray}
Further, by plugging Eqs. (\ref{9j}), (\ref{9k}) and (\ref{9m}) into (\ref{9g}), one can determine the magnetic coupling  parameter $\eta(r;L,\zeta,\alpha,\beta)$ as:
\begin{eqnarray}\label{9n}
\eta(r;L,\zeta,\alpha,\beta)=\sqrt{\frac{1}{U(r)}-\frac{L^2}{r^2 \zeta ^2}} \left(-\frac{\zeta ^2}{U(r)}+\frac{L^2}{r^2}+1\right).~~~~
\end{eqnarray}
Equation (\ref{9n}) indicates that a magnetized particle characterized by a magnetic coupling parameter $\eta$ is associated with a stable circular orbit that has angular momentum $L$ and energy $\zeta$.

Fig. (\ref{F4}) illustrates the magnetic coupling parameter for different values of $\alpha$ and $\beta$. We notice an upward trend in the parameter as both $\alpha$ and $\beta$ increase, while the parameter $\eta$ moves away from the singularity. It can be seen that as $\alpha$ and $\beta$ increase, $\eta$ also increases; however, at $ r = 4.3 $, all the curves for $\eta$ converge and subsequently begin to decline. For ISCOs, where the value of the parameter $\eta$ is:
\begin{eqnarray}\label{9o}
\eta=\eta(r;L,\zeta,\alpha,\beta),~~\frac{\partial \eta(r;L,\zeta,\alpha,\beta)}{\partial r}.
\end{eqnarray}
Eq. (\ref{9o}) consists of two equations featuring six unknown quantities. Given that there are five independent variables, the solution can be expressed in terms of two of these variables. For simplicity, we choose the radius $ r $ and $ \eta $ as free parameters.
The subsequent step involves showing how the specific energy $\zeta$ and the angular momentum $L$ depend on $r$. By applying the second condition from Eq. (\ref{9o}), one can initially determine the particle's minimum energy, which corresponds to the lowest value of the magnetic interaction. This information can then be used to calculate the specific energy.

Fig. (\ref{F4}) illustrates the values of the magnetic coupling parameters corresponding to various values of $\alpha$ and $\beta$, while maintaining a constant $ L = 5 $ and a specific energy of $ \zeta = 1 $. It is evident that reducing $\alpha$ and increasing $\beta$ leads to an increase in quantity $\eta$. Additionally, we note that as $\alpha$ increases and $\beta$ falls, the minimum value of $\eta$ ($\eta_{min}$) moves closer to the singularity.

To find the minimum value of $ L $ that corresponds to $ \eta_{min} $, the following condition must be met. This is expressed as:
\begin{small}
\begin{eqnarray}\label{9p}
&&\hspace{-0.5cm}\zeta_{min}=\frac{\sqrt{3 L^2+r^2} \sqrt{-\alpha  M-\alpha  M \log \left(\frac{r}{\beta }\right)+2 M r-r^2}}{\sqrt{3} r^2},
\\\label{9q}
&&\hspace{-0.5cm}\eta_{min}=\frac{2 \left(3 L^2+2 r^2\right) \sqrt{\frac{r^2 \left(6 L^2+r^2\right)}{\left(3 L^2+r^2\right) \left(\alpha  M \log \left(\frac{r}{\beta }\right)+M (\alpha -2 r)+r^2\right)}}}{3 r^2}.~~~
\end{eqnarray}
\end{small}
Fig. (\ref{F5}) illustrates the minimum specific energy of the magnetized particle for various values of $\alpha$ and $\beta$. The highest specific energy occurs when the parameter $\alpha$ decreases and $\beta$ increases. It is observed that as $\alpha$ increases and $\beta$ decreases, $\zeta_{min}$ moves closer to the singularity. By utilizing Eqs. (\ref{9o}) and (\ref{9p}), the expression for the minimum value can be derived as follows.
\begin{widetext}
\begin{align}\label{9r}
L_{min}=\frac{\sqrt{\left(6 M r^3-12 \alpha  M r^2 \log \left(\frac{r}{\beta }\right)-3 \alpha  M r^2+6 r^4\right)^2-4 \left(-18 \alpha  M-36 \alpha  M \log \left(\frac{r}{\beta }\right)+36 M r\right) L_{m1} (r)}}{2 \left(-18 \alpha  M-36 \alpha  M \log \left(\frac{r}{\beta }\right)+36 M r\right)},~~~~
\end{align}
\end{widetext}
where:
\begin{eqnarray}
    L_{m1} (r)=\left(2 M r^5-2 \alpha  M r^4 \log \left(\frac{r}{\beta }\right)-\alpha  M r^4\right). 
\end{eqnarray}

Fig. (\ref{F6}) illustrates the minimum value of the magnetic coupling parameter for different values of $\alpha$ and $\beta$, with a fixed value of $L = 5$ and a specific energy $\zeta = 1$. It is evident that as the values of $\alpha$ and $\beta$ increase, $\eta$ also increases. Additionally, we notice that as $\alpha$ and $\beta$ increase, $\eta_{min}$ moves away from the singularity. Fig. (\ref{F7}) shows the minimum specific angular momentum for different values of $\alpha$ and $\beta$. It is noted that as the values of $\alpha$ increase and the values of $\beta$ decrease, the specific angular momentum $ L_{min} $ also increases. Additionally, $ L_{min} $ tends to move towards the singularity as the value of $ \alpha $ decreases while considering all values of $ \beta $. By substituting Eq. (\ref{9r}) into Eq. (\ref{9q}), we can determine the extreme value of the magnetic coupling parameter $ \eta $:
\begin{widetext}
\begin{eqnarray}\label{9s}
&& \hspace{-0.5cm} \eta_{ext}= \frac{1}{36 \sqrt{2} M^2 \left(\alpha +2 \alpha  \log \left(\frac{r}{\beta }\right)-2 r\right)^2}\Bigg[\Big(-M^2 \left(7 r^2-96\right) (\alpha -2 r)^2+4 M r^4 (2 r-\alpha )\nonumber\\&&\hspace{0.5cm} +8 \alpha  M \log \left(\frac{r}{\beta }\right) \left(-2 \alpha  M \left(r^2-24\right) \log \left(\frac{r}{\beta }\right)+3 M \left(r^2-16\right) (2 r-\alpha )-2 r^4\right)+4 r^6\Big) \nonumber\\&&\hspace{0.5cm} \times \sqrt{-\frac{r^2 \left(-M^2 \left(7 r^2-24\right) (\alpha -2 r)^2+4 M r^4 (2 r-\alpha )+8 \alpha  M \log \left(\frac{r}{\beta }\right) \eta_1(r)+4 r^6\right)}{\left(\alpha  M \log \left(\frac{r}{\beta }\right)+M (\alpha -2 r)+r^2\right) \eta_3(r)}} \Bigg],~~~~~~~~
\end{eqnarray}
where:
\begin{align}
&    \eta_1(r)=\left(-2 \alpha  M \left(r^2-6\right) \log \left(\frac{r}{\beta }\right)+3 M \left(r^2-4\right) (2 r-\alpha )-2 r^4\right), \\
&    \eta_2(r)=8 \alpha  M \log \left(\frac{r}{\beta }\right)\left(2 \alpha  M \left(r^2-12\right) \log \left(\frac{r}{\beta }\right)-3 M \left(r^2-8\right) (2 r-\alpha )+2 r^4\right),\\
& \eta_3(r)=\left(M^2 \left(7 r^2-48\right) (\alpha -2 r)^2+4 M r^4 (\alpha -2 r)+ \eta_2(r)-4 r^6\right). 
\end{align}
\end{widetext}

Fig. (\ref{F8}) shows the maximum value of $\eta$ for various values of $\alpha$ and $\beta$. It is evident that $\eta_{ext}$ increases with higher values of $\alpha$ and $\beta$. As the radial coordinate increases, $\eta_{ext}$ also increases and approaches $1$ as $r \rightarrow \infty$. This indicates that for magnetized particles with $\eta_{ext} = \eta > 1$ in the vicinity of BHs surrounded by a magnetic field, stable circular orbits do not exist. Therefore, we can deduce $1 > \eta > \eta_{ext}$.
\begin{figure*}
\centering \epsfig{file=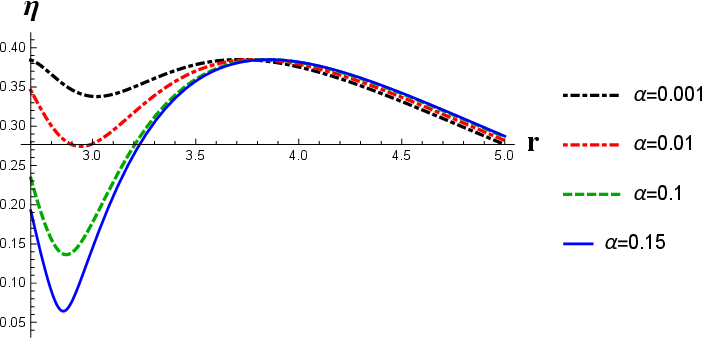, width=.4\linewidth,
height=1.7in}~~~~~~\epsfig{file=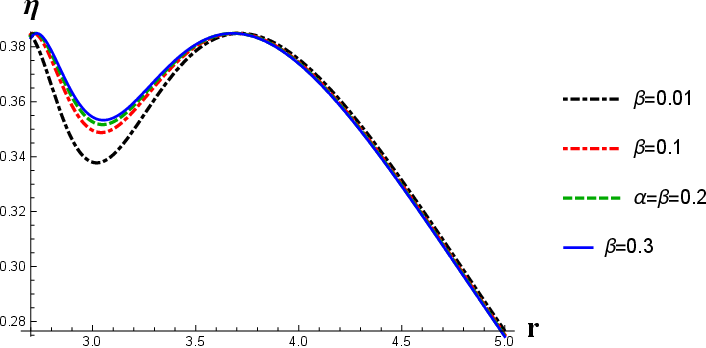, width=.4\linewidth,
height=1.7in} \caption{\label{F4} Variation in Magnetic coupling parameter $\eta$ is due to the variation of decoupled parameters $\alpha$ and $\beta$.
$\eta$ increased away from the singularity for increasing the parameters $\alpha$ and $\beta$.}
\end{figure*}
\begin{figure*}
\centering \epsfig{file=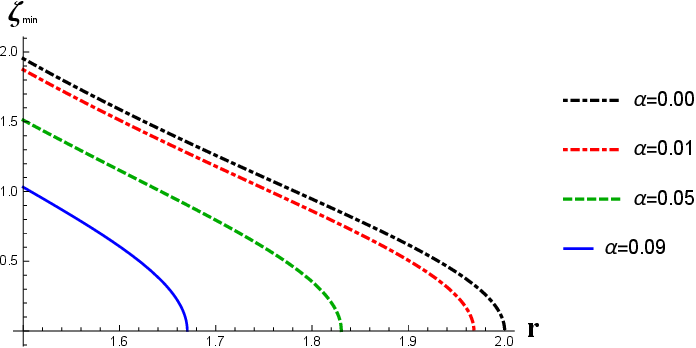, width=.4\linewidth,
height=1.7in}~~~~~~\epsfig{file=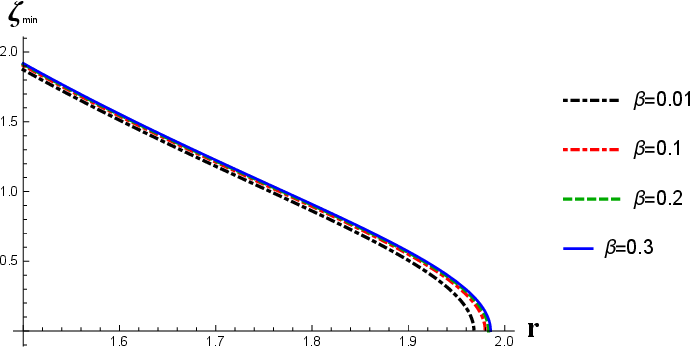, width=.4\linewidth,
height=1.7in} \caption{\label{F5} The changes in the specific energy of the magnetized particle parameter $\zeta$ are attributed to variations in the decoupled parameters $\alpha$ and $\beta$. Near the singularity, $\eta$ decreases as the parameter $\alpha$ increases, while it rises when moving away from the singularity.}
\end{figure*}
\begin{figure*}
\centering \epsfig{file=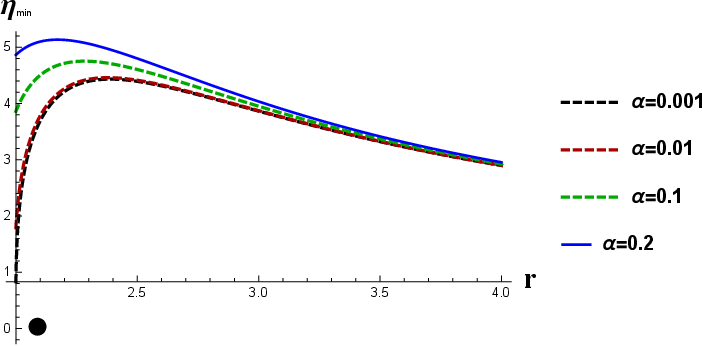, width=.4\linewidth,
height=2.02in}~~~~~~\epsfig{file=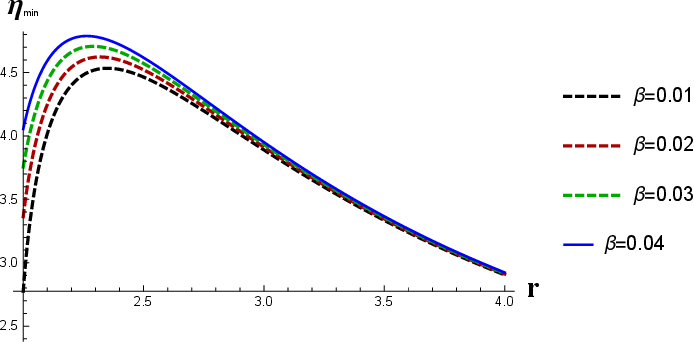, width=.4\linewidth,
height=2.02in}\caption{\label{F6} The changes in the minimal value of the magnetic coupling parameter $\eta$ are attributed to variations in the decoupled parameters $\alpha$ and $\beta$. As the parameters $\alpha$ and $\beta$ increase, $\eta_{\text{min}}$ rises as it moves away from the singularity, whereas it decreases as it approaches the singularity.}
\end{figure*}

\section{Circular motion around decoupled BH}

Using the standard conditions, the stable circular orbits of a magnetized particle in proximity to the compact object can be characterized as \cite{63f}:
\begin{eqnarray}\label{9t}
\dot{r}=0, \quad \frac{d}{dr}V_{eff}=0=\frac{d^2}{dr^2}V_{eff},
\end{eqnarray}
where $ V_{\text{eff}} $ represents the effective potential of the magnetized particle and is defined as:
\begin{eqnarray}
V_{eff}=U(r)\Big[\Big(1-\eta\sqrt{U(r)}\Big)^2+\frac{L^2}{r^2}\Big].\label{9u}
\end{eqnarray}
The effective potential is clearly influenced by the radial coordinate $r$, the angular momentum $ L $ and the metric function $U(r)$. When studying the motion of magnetized particles, the effective potential $ V_{eff} $ proves to be quite valuable. Specifically, the local extrema of $ V_{eff} $ are utilized to determine the positions of circular orbits.

Fig. (\ref{F9}) illustrates the radial profile of $ V_{eff} $ for different values of $ \alpha $ and $ \beta $. It is observed that $ V_{eff} $ increases as the values of $ \alpha $ and $ \beta $ diminish. Additionally, when the decoupled parameter $ \alpha $ decreases and $ \beta $ increases, $ V_{eff} $ moves closer to the singularity. By applying the second condition from Eq. (\ref{9t}), we can now compute $ L $ and $ \zeta $ for the magnetized particle in circular motion:
\begin{widetext}
\begin{eqnarray}
L^2&=&\sqrt{\frac{U''(r) r^3 \left(2 \eta ^2 -3 \eta  \sqrt{U(r)}+1\right)}{2 U(r)-U''(r)r}},\label{9v}\\
\label{9w}
\zeta^2&=&\sqrt{\frac{2 \eta ^2 U(r)^{3/2}-4 \eta  U(r)+\eta ^2 \sqrt{U(r)} U''(r) r+2 \sqrt{U(r)}-\eta  U''(r) r}{U(r)^{-3/2} (2 U(r)-U''(r) r)}}.
\end{eqnarray}
\end{widetext}
We analyze how the parameters $\alpha$ and $\beta$ influence the quantities $L$ and $\zeta$ of magnetized particles, specifically with $\eta = 1$, as depicted in Figs. (\ref{F10}) and (\ref{F11}). In Fig. (\ref{F10}), we present the radial profiles of $L$ for test particles in the vicinity of a decoupled BH, considering various values of $\alpha$ and $\beta$. The results indicate that an increase in the values of $\alpha$ and $\beta$ leads to a corresponding increase in $L$. Additionally, we note that as the parameters $\alpha$ and $\beta$ increase, the angular momentum $L$ moves away from the singularity. Fig. (\ref{F11}) illustrates the specific energy of test particles near a decoupled BH for various values of $\alpha$ and $\beta$. It is evident that $\zeta$ increases with an increase in $\alpha$ and a decrease in $\beta$. Furthermore, we observe that when $\alpha$ decreases and $\beta$ increases, $\zeta$ moves closer to the singularity.

\begin{figure*}
\centering \epsfig{file=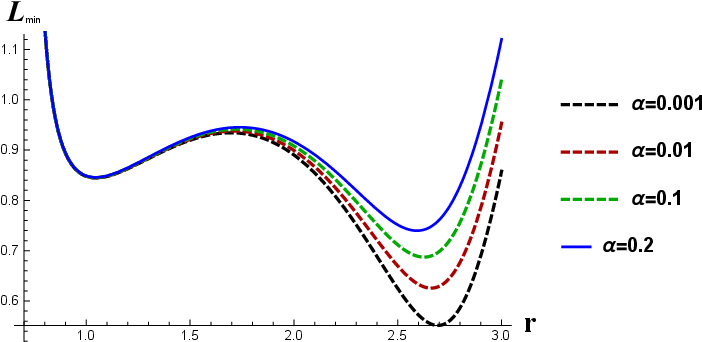, width=.45\linewidth,
height=2in}~~~~~~\epsfig{file=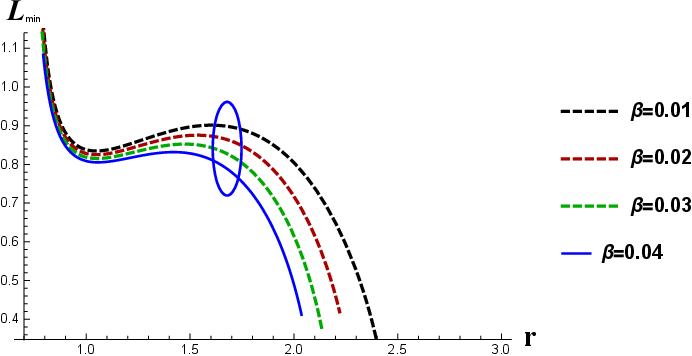, width=.45\linewidth,
height=2.in}\caption{\label{F7} The minimum specific angular momentum model arises from changes in the decoupled parameters $\alpha$ and $\beta$. The value of $L_{min}$ increases as the parameter $\alpha$ decreases and the parameter $\beta$ increases, moving away from the singularity. Conversely, $L_{min}$ decreases towards the singularity when the parameter $\alpha$ is reduced.}
\end{figure*}

\begin{figure*}
\centering \epsfig{file=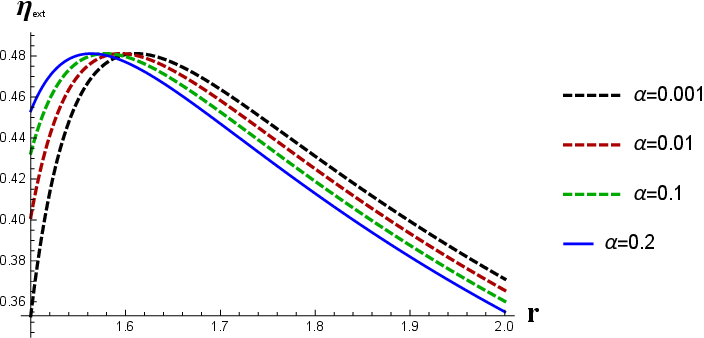, width=.45\linewidth,
height=2.2in}~~~~~~\epsfig{file=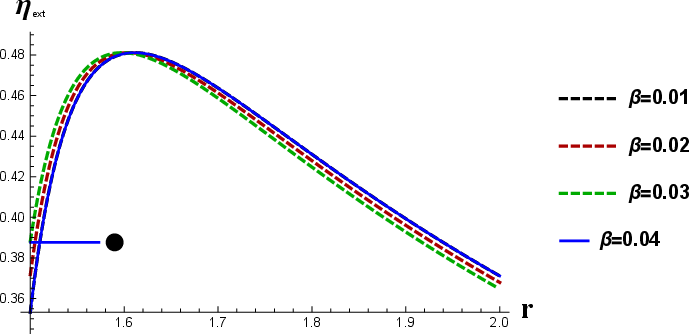, width=.45\linewidth,
height=2.02in}\caption{\label{F8} The extreme value of the $\eta$ model results from changes in the decoupled parameters $\alpha$ and $\beta$. As the parameters $\alpha$ and $\beta$ increase, $\eta_{ext}$ rises away from the singularity.}
\end{figure*}

\begin{figure*}
\centering \epsfig{file=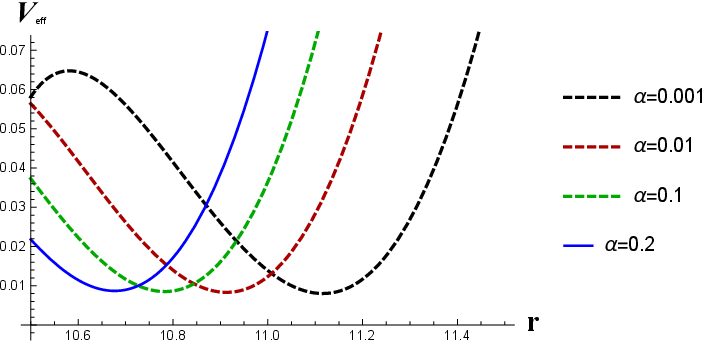, width=.45\linewidth,
height=2in}~~~~~~\epsfig{file=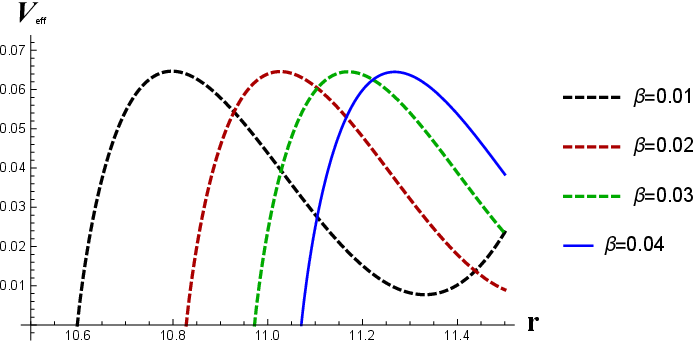, width=.45\linewidth,
height=1.95in}\caption{\label{F9} The effective potential for magnetized particles arises from changes in the decoupled parameters $\alpha$ and $\beta$. The effective potential $V_{eff}$ increases as the parameter $\alpha$ decreases and $\beta$ increases, moving away from the singularity.}
\end{figure*}

\begin{figure*}
\centering \epsfig{file=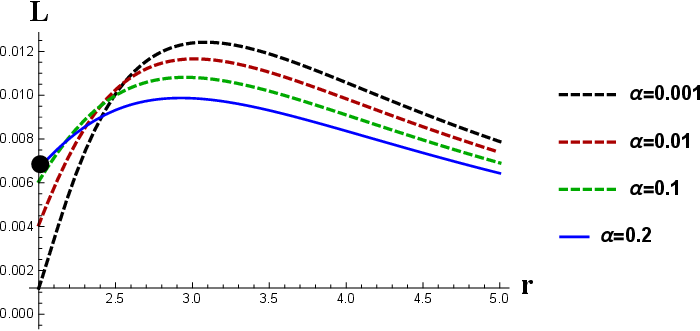, width=.45\linewidth,
height=2.02in}~~~~~~\epsfig{file=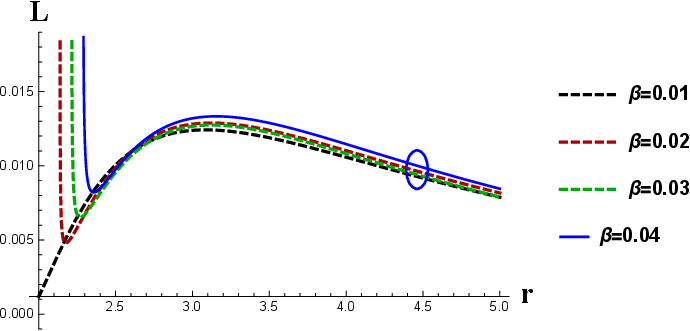, width=.45\linewidth,
height=2.02in}\caption{\label{F10} The radial profiles of $ L $ for test particles near the decoupled parameters $ \alpha $ and $ \beta $ show that these parameters influence its variation. As the parameters $ \alpha $ and $ \beta $ increase, the value of $ L $ rises as it moves away from the singularity.}
\end{figure*}

\subsection{Effective force on decoupled BH}

If a particle is drawn towards or is moving away from the BH, the effective potential $ V_{eff} $ that influences the particle provides insights into its behavior. In this context, we utilize Eq. (\ref{9u}) to determine the effective force acting on the particle as follows:
\begin{eqnarray}
F_{eff}&=&\frac{-1}{2}\frac{d}{dr}V_{eff}.\label{9x}
\end{eqnarray}
The radial profile of the effective force $ F_{eff} $ is illustrated in Fig. (\ref{F12}) for various values of $ \alpha $ and $ \beta $. It is observed that $ F_{eff} $ increases with increasing values of both $ \alpha $ and $ \beta $.

\begin{figure*}
\centering \epsfig{file=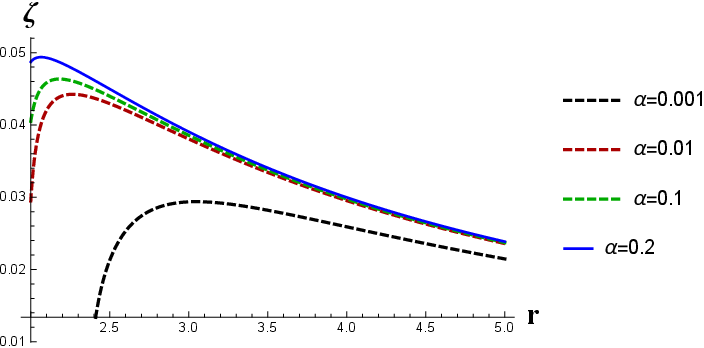, width=.45\linewidth,
height=2in}~~~~~~\epsfig{file=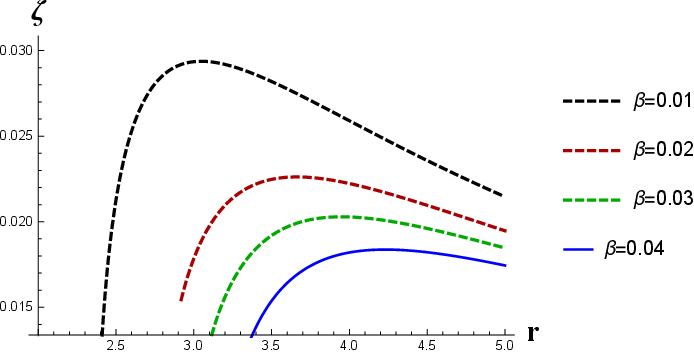, width=.45\linewidth,
height=1.95in}\caption{\label{F11} The specific energy for test particles near a decoupled BH varies based on the parameters $\alpha$ and $\beta$. As the parameter $\alpha$ increases and the parameter $\beta$ decreases, $\zeta$ also increases. The other parameters are set to $M=1$ and $\eta=0.01$.}
\end{figure*}

\begin{figure*}
\centering \epsfig{file=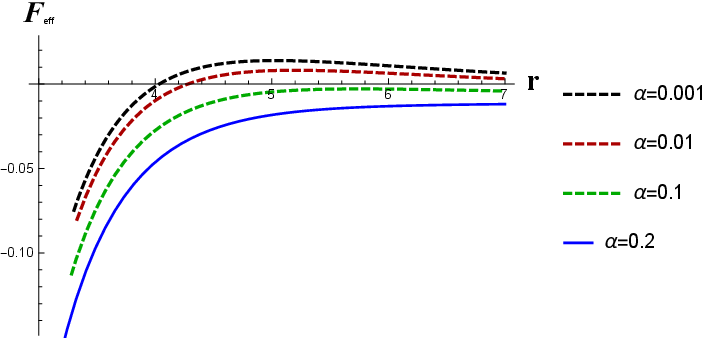, width=.45\linewidth,
height=2in}~~~~~~\epsfig{file=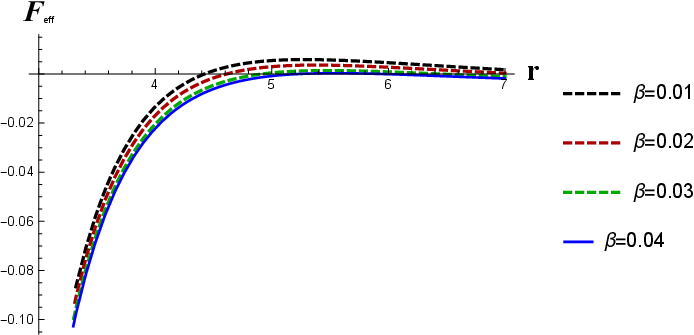, width=.45\linewidth,
height=2in}\caption{\label{F12} The effective force $ F_{eff} $ arises from the changes in the decoupled parameters $ \alpha $ and $ \beta $. As the values of $ \alpha $ and $ \beta $ increase, $ F_{eff} $ becomes larger as it moves away from the singularity.}
\end{figure*}

\begin{figure*}
\centering \epsfig{file=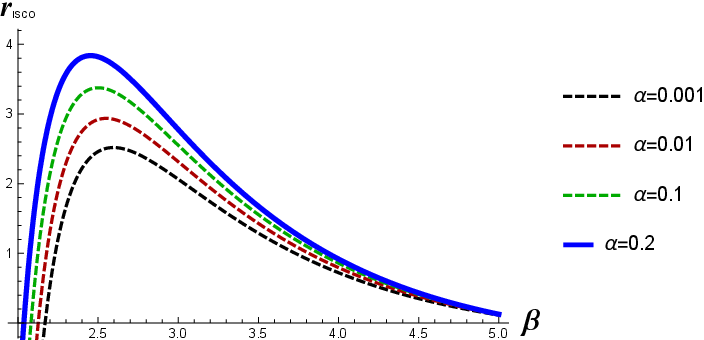, width=.45\linewidth,
height=2in}~~~~~~\epsfig{file=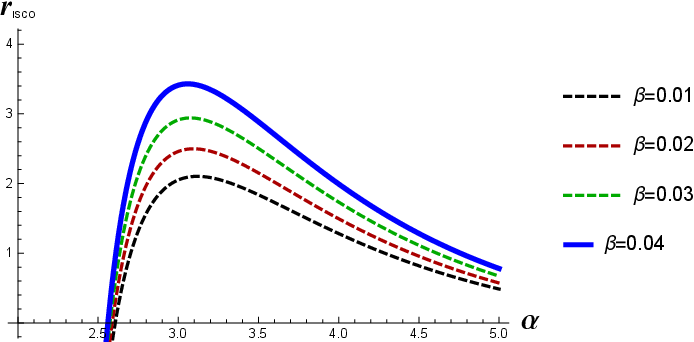, width=.45\linewidth,
height=2in}
\caption{\label{F13} The ISCOs radius model arises from the changes in the decoupled parameters $\alpha$ and $\beta$. For all values of $\alpha$ and $\beta$, the ISCO radius rapidly approaches the singularity.}
\end{figure*}

\begin{figure*}
\begin{center}
\includegraphics[width=160mm]{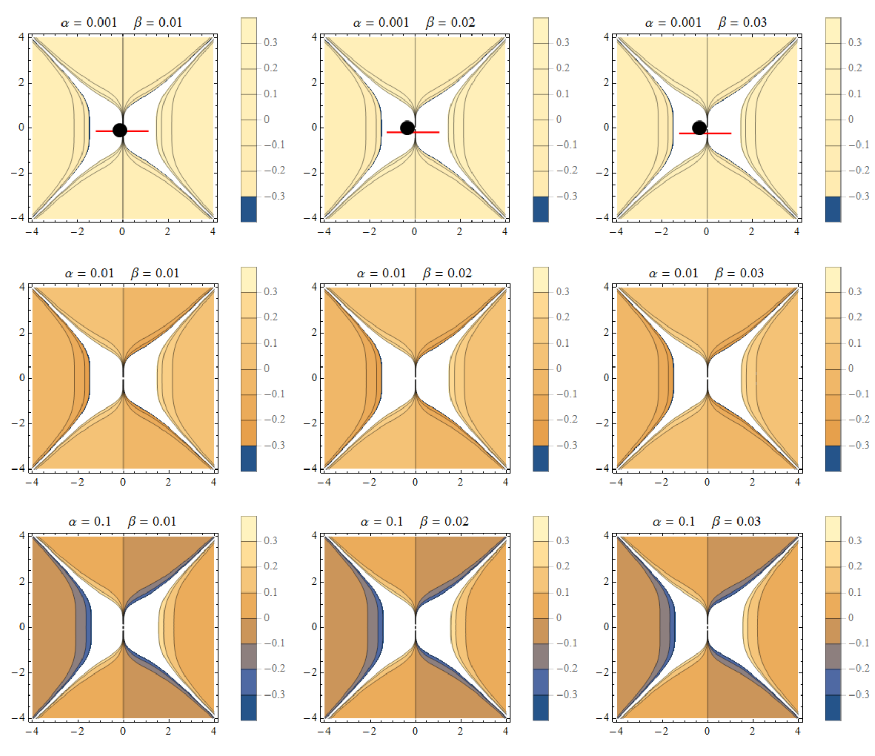}
\caption{\label{F14} The BH model describes the movement of a  particle using the decoupled parameters $\alpha$ and $\beta$.}
\end{center}
\end{figure*}

\begin{figure*}
\begin{center}
\includegraphics[width=160mm]{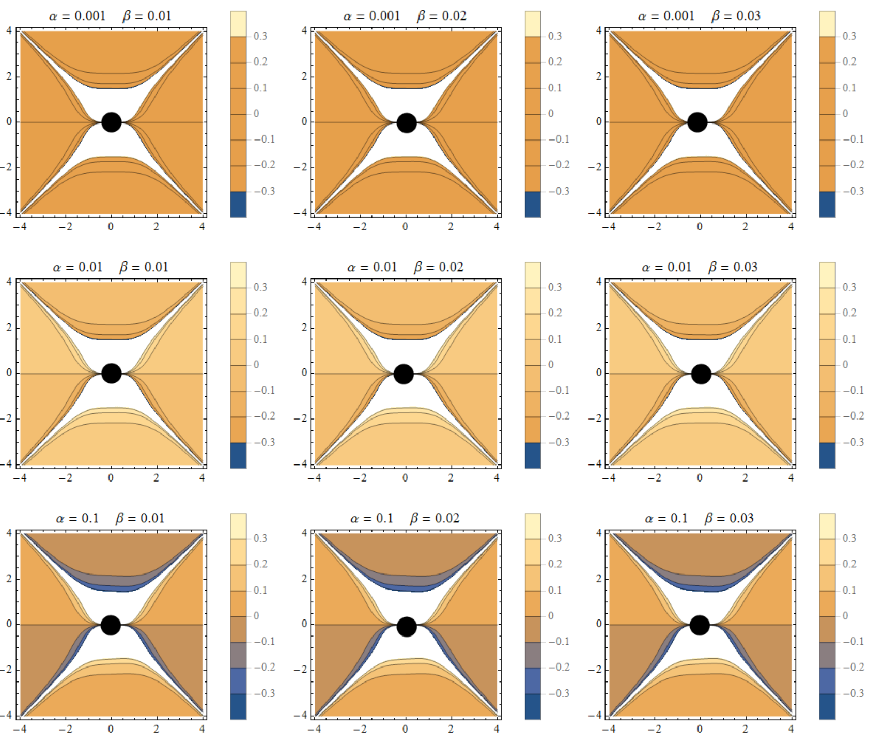}
\caption{\label{F15} The BH model describes the motion of a particle using the decoupled parameters $\alpha$ and $\beta$.}
\end{center}
\end{figure*}

\subsection{ISCOs radius of decoupled BH}

Once again, taking Eq. (\ref{9t}), we can determine the radius of ISCOs by using the following condition:

\begin{eqnarray}\label{9y}
\frac{d^2 V_{eff}}{dr^2}=0,
\end{eqnarray}
An exact analytical solution for Eq. (\ref{9y}) is clearly unattainable. Therefore, we conduct a comprehensive numerical analysis in order to find the ISCOs. Fig.(\ref{F13}) illustrates how the radius of the ISCOs varies with $\alpha$ and $\beta$. It is evident that as $\alpha$ increases and $\beta$ decreases, the radius of the ISCOs moves closer to the singularity. Secondly, once can notice from the same Fig.(\ref{F13}) that numerically calculated ISCOs has a good agreement with the standard model like a Schwarzschild BH.

\subsection{Trajectories of magnetized particles around decoupled BH}

This section analyzes the bound orbits of magnetized particles in the vicinity of a decoupled BH exposed to an external magnetic field. The particle's angular momentum is fixed at $L = 5$ with $\eta = 1$. The trajectories of the magnetized particles for different values of the parameters $\alpha$ and $\beta$ are illustrated in Figs. (\ref{F14}) and (\ref{F15}). The presence of parameters $\alpha$ and $\beta$ is observed to lead to an increase in the radius of the bound orbits and the energy of the particles along these orbits. The analysis concludes that the gravitational potential of the decoupled BH increases for both lower and higher values of $\alpha$ and $\beta$.

\section{Basic parameters of accretion process around decoupled BH} \label{se5}

In this section, we present some fundamental calculations of accretion as derived by Babichev et al. \cite{gmr2,gmr3,gmr4}. To do this, we consider the energy-momentum tensor of a perfect fluid, which is expressed as follows:
\begin{eqnarray}
T^{\mu\nu} =(\rho+p)u^\mu u^\nu + pg^ {\mu \nu},\label{32}
\end{eqnarray}
where $\rho$ and $p$ represent the energy density and pressure of the accreting fluid, respectively, while $u^\mu$ denotes the four-velocity of the fluid. In the equatorial plane, where $\theta=\pi/2$, the four-velocity can be expressed in its general form as:
\begin{eqnarray}
u^\mu = \frac{dx^\mu}{d\tau}=(u^t,u^r,0,0),\label{33}
\end{eqnarray}
where $\tau$ represents the appropriate time associated with the geodesic movement of the particles. The flow, which is stable and symmetric, satisfies the normalization condition $ u^\mu u_\mu = -1 $. From this we can derive the following:
\begin{eqnarray}
u^t =\frac{\sqrt{U(r)+(u^r)^2 }}{U(r)}.\label{34}
\end{eqnarray}
Due to square root, both $ u^t $ and $ u^r $ can have positive or negative values, indicating whether the conditions are for forward or backward time. In the case of $ u^r < 0 $ (inward flow), accretion occurs and the fluid's velocity is negative. In contrast, when $ u^r > 0 $ (outflow), the velocity of the fluid is positive. Thus, it is essential to consider the conservation laws of energy and momentum when analyzing accretion. The conservation of energy can be expressed as follows:
\begin{eqnarray}
T^{\mu\nu}_{;\mu} =0 \quad\Rightarrow\quad T^{\mu\nu}_{;\mu}=\frac{1}{\sqrt{-g}}(\sqrt{-g}T^{\mu\nu})_{,\mu}+\Gamma^{\nu}_{\alpha\mu}T^{\alpha\mu}=0,\label{35}
\end{eqnarray}
where $;$ denotes the covariant derivative, and $\Gamma^{\nu}_{\alpha\mu}$ represents the Christoffel's symbol of the second kind. After making simplifications, we arrive at:
\begin{eqnarray}
r^2u^r(\rho+p)\sqrt{U(r)+(u^r)^2}=N_0,\label{36}
\end{eqnarray}
where $N_0$ represents the integration constant.
From the relation between conservation law and 4-velocity
via $u_{\mu}T^{\mu\nu}_{;\nu}=0$, we calculate:
\begin{eqnarray}
(\rho+p)_{;\nu}u_{\mu}u^{\mu}u^{\nu}+(\rho+p)u^{\mu}_{;\nu}u_{\mu}u^{\nu}+(\rho+p)u_{\mu}u^{\mu}u^{\nu}_{;\nu}\nonumber\\+p_{,\nu}g^{\mu\nu}u_{\mu}+p u_{\mu}g^{\mu\nu}_{;\nu}=0.~~\label{37}
\end{eqnarray}
By using the conditions $ u^\mu u_\mu = -1$ and $g^{\mu\nu}_{;\nu}=0$, the above equation reduces to:
\begin{eqnarray}
(\rho+p)u^{\nu}_{;\nu}+u^{\nu}\rho_{\nu}=0.\label{38}
\end{eqnarray}
Taking only the non-zero components, we get:
\begin{eqnarray}
\frac{\rho'}{\rho+p}+\frac{u'}{u}+\frac{2r}{r^2}=0.\label{39}
\end{eqnarray}
By integrating the above equation, we are left with:
\begin{eqnarray}
r^2u^{r}\exp\int\frac{d\rho}{\rho+p}=-N_{1},\label{40}
\end{eqnarray}
where $N_{1}$ represents an integration constant.
Using $u^r<0$, the minus sign is taken on the right-hand side, so we get the following:
\begin{eqnarray}
(\rho+p)\sqrt{\left[(u^r)^2+f(r)\right]}\exp\left(-\int\frac{d\rho}{\rho+p}\right)=N_{2},\label{41}
\end{eqnarray}
with $N_{2}$ being a constant of integration.
By using the above setup, the mass flux is given by:
\begin{eqnarray}
(\rho u^\mu)_{;\mu}\equiv\frac{1}{\sqrt{-g}}(\sqrt{-g}\rho u^\mu)_{,\mu}=0,\label{42}
\end{eqnarray}
which can also be written as:
\begin{eqnarray}
\frac{1}{\sqrt{-g}}(\sqrt{-g}\rho u^\mu)_{,r}+\frac{1}{\sqrt{-g}}(\sqrt{-g}\rho u^\mu)_{,\theta}=0.\label{43}
\end{eqnarray}
Therefore, the conservation mass equation is given by:
\begin{eqnarray}
r^2\rho u^r=N_{3},\label{44}
\end{eqnarray}
where $N_{3}$ symbolizes the integration constant.

\subsection{Radial velocity and energy density profiles for decoupled BH}

To proceed, we consider an isothermal fluid characterized by the equation of state $ p = \omega \rho $, where $ \omega $ is the state parameter. In these fluids, the flow must maintain a constant temperature. During the accretion process, the speed of sound in such fluids remains constant, following the relationship $ p \propto \rho $. Consequently, from Eqs. (\ref{40}), (\ref{41}), and (\ref{44}), we obtain the following results:
\begin{eqnarray}
\left(\frac{\rho+p}{\rho}\right)\sqrt{\left[(u^r)^2+U(r)\right]}\exp\left(-\int\frac{d\rho}{\rho+p}\right)=N_{4},\label{45}
\end{eqnarray}
where $N_{4}$ is the integration constant. Using $p=\omega\rho$ in the above equation, we get:
\begin{eqnarray}
u(r)=\left(\frac{1}{k+1}\right)\sqrt{\frac{(N_{4})^2}{U(r)}-(k+1)^2}.\label{46}
\end{eqnarray}
Therefore, the radial velocity of strong and weak fields is given by:
\begin{eqnarray}\label{47}
u(r)&=&\left(\frac{1}{k+1}\right)\sqrt{\frac{(N_{4})^2}{\left(\ln \left(\frac{r}{\beta }\right)\frac{(\alpha  M)}{r^2} +\frac{\alpha  M}{r^2}-\frac{2 M}{r}+1\right)}
-(k+1)^2}.
\end{eqnarray}
From Eq. (\ref{44}), we obtain the density of the fluid expressed as follows:
\begin{eqnarray}\label{49}
\rho(r)=\frac{N_{3}}{r^2}\frac{(k+1)}{\sqrt{{\frac{(N_{4})^2}{U(r)}-(k+1)^2}}}.
\end{eqnarray}
Therefore, the energy density of strong and weak fields becomes:
\begin{eqnarray}\label{50}
\rho(r)&=&\frac{N_{3}}{r^2}\frac{(k+1)}{\sqrt{{\frac{(N_{4})^2}{\left(\ln \left(\frac{r}{\beta }\right)\frac{(\alpha  M)}{r^2} +\frac{\alpha  M}{r^2}-\frac{2 M}{r}+1\right)}-(k+1)^2}}}.~~~
\end{eqnarray}

\begin{figure*}
\centering \epsfig{file=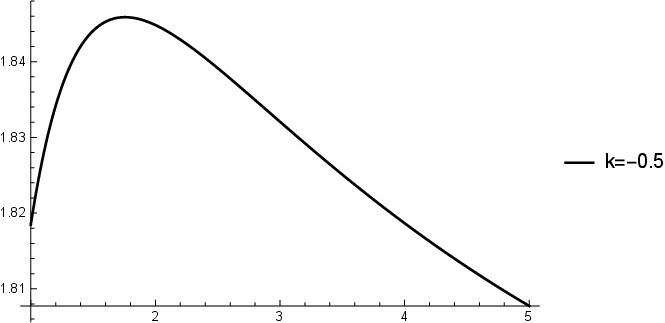, width=2.6in,height=1.7in}~~~~~~~~~\epsfig{file=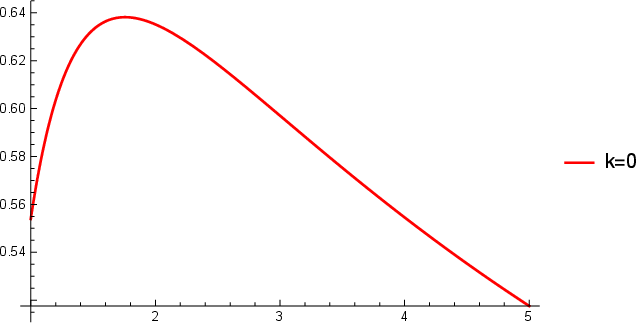, width=2.6in,height=1.7in}
\vspace{0.5cm}
\centering \epsfig{file=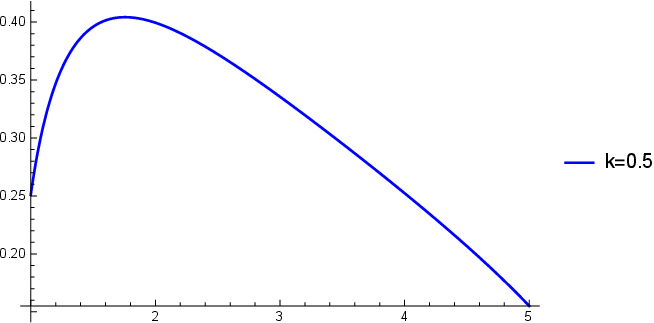, width=2.6in,height=1.7in}~~~~~~~~~\epsfig{file=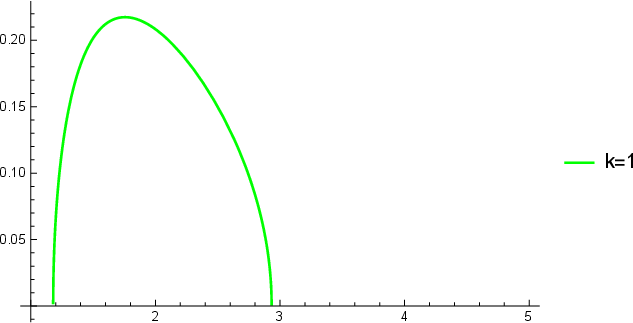, width=2.6in,height=1.7in}
\vspace{0.5cm}
\centering \epsfig{file=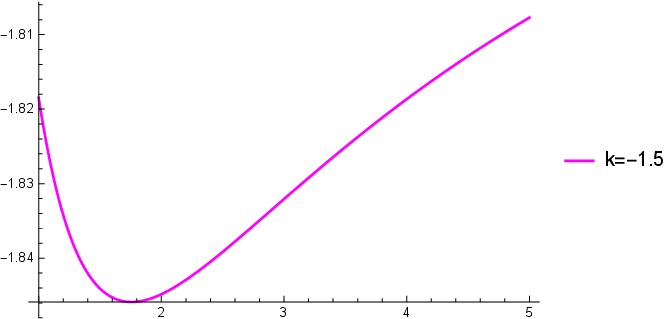, width=2.6in,height=1.7in}~~~~~~~~~\epsfig{file=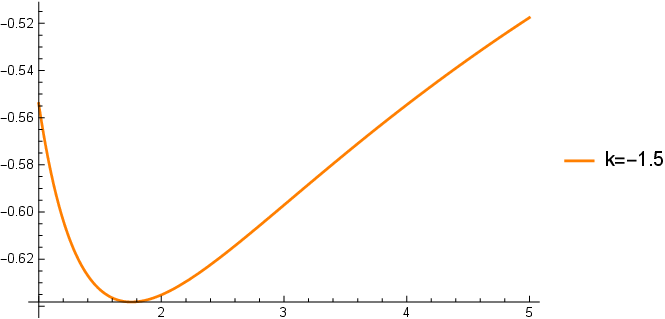, width=2.6in,height=1.7in}
\caption{\label{F16} The radial velocity profile of decoupled BHs represents the four accretion scenarios. The changes in the state parameter $ k $ are depicted in the plotted scenarios.}

\end{figure*}
\begin{figure*}
\centering \epsfig{file=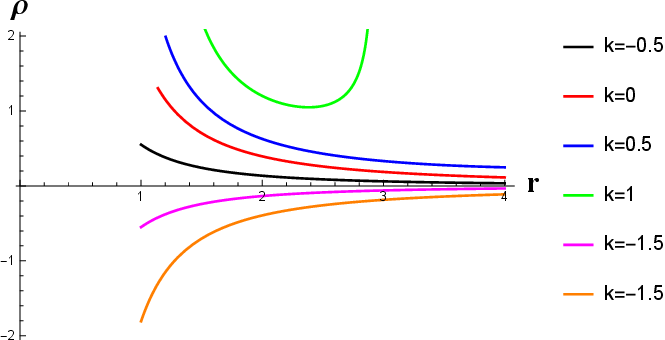, width=.45\linewidth,
height=2in}~~~~~~~~~\epsfig{file=rou.eps, width=.45\linewidth,
height=2in}\\
\vspace{0.5cm}
\centering \epsfig{file=rou.eps, width=.45\linewidth,
height=2in}~~~~~~~~~\epsfig{file=rou.eps, width=.45\linewidth,
height=2in}\caption{\label{F17} The energy density profile of decoupled BHs illustrates the four different accretion scenarios. The changes in the state parameter $ k $ are shown in the plotted scenarios.}
\end{figure*}

\begin{figure*}
\centering \epsfig{file=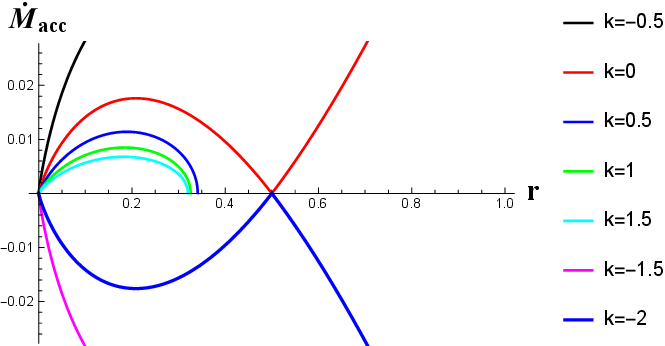, width=.43\linewidth,
height=2in}~~~~~~\epsfig{file=accretion.eps, width=.43\linewidth,
height=2in}\\
\vspace{0.5cm}
\centering \epsfig{file=accretion.eps, width=.43\linewidth,
height=2in}~~~~~~~\epsfig{file=accretion.eps, width=.43\linewidth,
height=2in}
\caption{\label{F18} The change in mass of decoupled BHs reflects four different accretion scenarios. The changes in the state parameter $ k $ are depicted in the scenario plots.}
\end{figure*}

\subsection{Sonic point accretion around decoupled BH}

As the fluid element remains stationary far from the BH while moving inward, it must traverse a critical point where the fluid's velocity matches the speed of sound. Maximum accretion occurs when the fluid is directed towards this critical point. Considering that $ h = h(\rho) $ represents a constant enthalpy, the fluid behaves in a barotropic manner. Consequently, the equation of state is provided by \cite{57f}:
\begin{eqnarray}
\frac{dh}{h} = V^2\frac{d\rho}{\rho},\label{51}
\end{eqnarray}
where $V$ is the local speed of sound. This equation further gives $\ln h=V^2\ln n$.
So, from Eqs. (\ref{44}), (\ref{45}) and (\ref{51}), we obtain:
\begin{eqnarray}
\left[\left(\frac{u}{u_t}\right)^2-V^2\right](\ln u)_{,r}=\frac{1}{r^2(u_t)}\left[2rV^2(u_t)^2-\frac{1}{2}r^2 U'(r)\right],\label{52}
\end{eqnarray}
where critical points are indicated by the subscripted letter $ c $, allowing determination of the solution for the local speed of sound at these points. This is expressed as:
\begin{eqnarray}
V^2_c = \left(\frac{u_c}{u_{tc}}\right)^2.\label{53}
\end{eqnarray}

At the sonic points, we have the following:
\begin{eqnarray}
2rV^2_c(u_{tc})^2-\frac{1}{2}r^2_c U'_{rc} = 0,\label{54}
\end{eqnarray}
where $U_c=U(r)|_{r=r_c}$ and $U'_{rc}=U'(r_c)$.
We obtain the radial velocity at the critical point by putting Eq. (\ref{53}) into (\ref{54}) given by:
\begin{eqnarray}
(u_c)^2 = \frac{1}{4r}r^2_c U'_{rc}.\label{55}
\end{eqnarray}
By using Eqs. (\ref{34}), (\ref{54}) and (\ref{55}), we obtain:
\begin{eqnarray}
r^2_c U'_{rc} = 4rV^2_c[U(r_c)+(u_c)^2],\label{56}
\end{eqnarray}
which produces the local speed of sound given as follows:
\begin{eqnarray}
V^2_c = \frac{r^2_c U'_{rc}}{r^2_c U'_{r_c}+4rU(r_c)}.\label{57}
\end{eqnarray}
Using Eq. (\ref{45}), the speed of sound is calculated as:
\begin{eqnarray}
c^2_s = \frac{N_4}{\sqrt{U(r)+u^2_c}}-1.\label{57a}
\end{eqnarray}
By using the Hamiltonian approach \cite{t14f,t15f}, we have:
\begin{eqnarray}
H=\frac{U^{1-k}(r)}{(1-\omega^2)^{1-k}\omega^{2k}r^{4k}},\label{57b}
\end{eqnarray}
where $\omega \equiv \frac{dr}{G(r)dt}$ represents the three-dimensional velocity of a particle undergoing radial motion in the equatorial plane, which is further specified by:
\begin{eqnarray}
\omega^2=\left(\frac{u}{U(r)u^0}\right)^2=\frac{u^2}{u^2_0}=\frac{u^2}{U(r)+u^2}.\label{57c}
\end{eqnarray}
The primary critical solutions are examined at the sonic points derived from Eqs. (\ref{55}) and (\ref{56}). They are given as:
\begin{eqnarray}
(u_c)^2&=&\frac{1}{4}r_c U'_c(r),\label{57d}\\
(u_c)^2 &=&k\left(U_c(r)+\frac{1}{4}r_c U'_c(r)\right).\label{57e}
\end{eqnarray}
This generalized result can be examined by selecting any value of $\omega$. In this study, we will explore various interesting fluids, including ultra-stiff fluid $(k=1)$, ultra-relativistic fluid $(k=1/2)$, radiation fluid $(k=1/3)$, and sub-relativistic fluid $(k=1/4)$ in the context of accretion flow surrounding a Weyl geometric BH.

\subsection{Area times flux for decoupled BH}

The mass accretion rate refers to how quickly the mass of a BH changes, which is determined by the area multiplied by the flux at the BH's boundary. Essentially, this rate indicates the mass gained by the BH over a specific time period. It is influenced by the characteristics of the fluid being accreted and the metric parameters involved. In the context of quintessence in astrophysical scenarios, the mass of the BH is not constant. Processes like fluid accretion onto the BH and Hawking radiation cause the mass to vary gradually. The change in accretion mass can be calculated by integrating the fluid flux in the vicinity of the BH, represented by $\dot{M}$. Thus, it is expressed as follows:
\begin{eqnarray}
\dot{M_{acc}}=-4\pi r^2 u^r(\rho+p)\sqrt{U(r)+(u^r)^2}\equiv-4\pi N_0,\label{52}
\end{eqnarray}
where $N_0=-N_1N_2$ and $N_2=(p_\infty+\rho\infty)\sqrt{U(r_\infty)}$. Substituting them into the above equation leads to the following:
\begin{eqnarray}
\dot{M_{acc}}=4\pi N_{1}(\rho_{\infty}+p{_\infty})\sqrt{U(r_{\infty})}M^2.\label{53}
\end{eqnarray}
Here, we define our boundary at infinity (i.e., $ r = r_{\infty} $). The radius at infinity refers to a specific point in infinity where a massive particle is falling into the BH. In addition, $ \rho_{\infty} $ represents the energy density in $ r_{\infty} $, and $ p_{\infty} $ denotes the pressure in $ r_{\infty} $. We consider the time evolution of the BH mass, allowing us to express the above equation in the following manner:
\begin{eqnarray}
\frac{dM}{M^2}=  \mathcal{F}dt,\label{54}
\end{eqnarray}
where $\mathcal{F}=4\pi N_{1}(\rho_{\infty}+p{_\infty})\sqrt{U(r_{\infty})}$. By integrating the above equation, we have the following:
\begin{eqnarray}
M_t=\frac{M_i}{1-Ft M_i}\equiv\frac{M_i}{1-\frac{t}{t_{cr}}}.\label{55}
\end{eqnarray}
Let $ M_i $ represent the mass of the BH in its initial state, while $ M_t $ denote the mass of the BH at the critical accretion time. The critical accretion time is defined as:
\begin{eqnarray*}
\left(t_{cr}=\left[4\pi N_{1}\left(\rho_{\infty}+p_{\infty}\right)\sqrt{f(r_{\infty})M_{i}}\right]^{-1}\right).
\end{eqnarray*} 
Consequently, the expression for the BH mass accretion rate can be expressed as follows:
\begin{eqnarray}
\dot{M_{acc}}=4\pi N_{1}(\rho+p)M^2.\label{56}
\end{eqnarray}

Fig. (\ref{F16}) presents the velocity profile as a function of the radial coordinate for various values of the state parameter $k$, along with the decoupled parameters $\alpha$ and $\beta$. We consider $k=0$ for dust and $k=1$ for stiff matter. Furthermore, $k=-1$ corresponds to the phantom energy, while $-1 < k < -1/3$ is associated with the quintessence energy. The fluid is observed to exhibit a negative radial velocity for $k=-1.5$ and $k=-2$, whereas it shows a positive radial velocity for $k=0$, $k=0.5$, $k=1$, and $k=1.5$. Fig. (\ref{F17}) shows the energy density in relation to the radial coordinate for different values of $k$ and the Weyl parameters $c_1$ and $c_2$. When $k = -1.5$ or $-2$, the energy density decreases, while for $k = 0, 0.5, 1, 1.5$, this factor increases. Fig. (\ref{F18}) illustrates the relationship between mass and radial coordinate for various values of $k$, along with the decoupled parameters $\alpha$ and $\beta$.

Fig. (\ref{F19}) showing the contribution of URF, RF and SRF in the form of an accretion disk for hairy BH along with the decoupled parameters $\alpha$ and $\beta$. We observe the following key points.
\begin{itemize}
  \item We observe that the starting point of the fluid
outflow is at the horizon due to its very high pressure, which influences
the divergence and as a result the fluid with its own pressure
flows back to spatial infinity \cite{58}. 
 \item We also
observe that the supersonic accretion followed by
subsonic accretion stops inside the horizon and does not support the claim
that ``the flow must be supersonic at the horizon'' \cite{58x}. This means that for Hairy BH, the flow of the fluid is neither supersonic nor transonic near the
horizon \cite{59,60}. These results agree with fine-tuning and instability issues in dynamical systems.
The stability issue is related to the nature of the saddle points of the Hamiltonian system. The analysis of
stability could be done using Lyapunov's theorem or linearization
of the dynamical system \cite{61,62,63} and their variations \cite{64}.

 \item Another stability issue is that the flow of the fluid
starts in the surrounding of horizon under the effect of divergent pressure. This outflow is unstable because it follows a subsonic
path passing through the saddle point  and becomes
supersonic with a speed approaching the speed of light. The point
($r=r_h, \omega=0$) can be observed as both an attractor and a repeller
where the solution curves converge and diverge from a
cosmological point of view \cite{58x,64}.
 
\end{itemize}

The mass of the Weyl geometric BH increases with the accumulation of dust, quintessence, and stiff matter, while it decreases with the accumulation of phantom-like fluids.

\begin{figure*}
\centering \epsfig{file=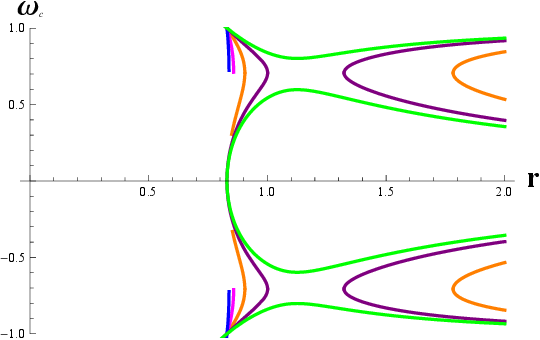, width=.35\linewidth,
height=2in}~~~~~~~~~~~\epsfig{file=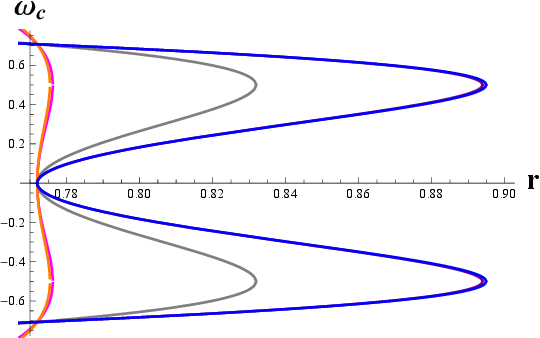, width=.35\linewidth,
height=2in}\\
\vspace{0.5cm}
\centering \epsfig{file=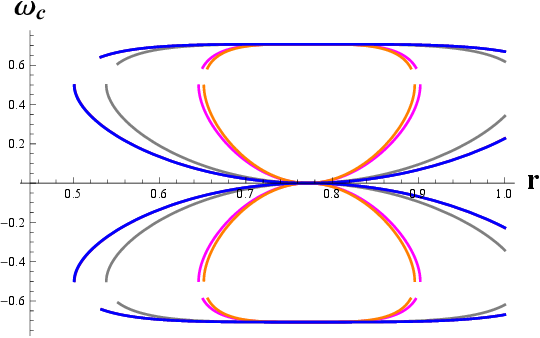, width=.35\linewidth,
height=2.0in}
\caption{\label{F19} The accretion patterns of the decoupled BH are shown based on the parameters $\alpha$ and $\beta$. The left plot illustrates the Ultra-Relativistic Flow (URF) as $H_c$ varies. The right plot depicts the Relativistic Flow (RF) with changes in $H_c$. The bottom plot represents the (SRF) in relation to the variation in $H_c$.}
\end{figure*}

\subsection{Electromagnetic emissivity of the thin accretion disks
around decoupled BH}

It is observed that materials falling from rest at infinity accumulate onto the BH. The gravitational energy released by these falling materials can convert into radiation, which is responsible for some of the most energetic astrophysical events. The energy flux of radiation across the accretion disk is related to the specific energy $\eta$, the angular momentum $L$ and the angular velocity $\Omega_\phi$ as analyzed by Kato et al. \cite{36dff}. Thus, we have:
\begin{eqnarray}
K=-\frac{\dot{M}\Omega_{\phi,r}}{4\pi \sqrt{-g}(\eta-L\Omega_\phi)^2}\int(\eta-L\Omega_\phi)L_{,r}dr,\label{22}
\end{eqnarray}
where $ K $ represents the radiation flux, $ \Omega_{\phi,r} = \frac{d\Omega_\phi}{dr} $, and $ \dot{M} $ denotes the accretion rate. Since our analysis is limited to the equatorial plane, we set $ \sin\theta = 1 $. Based on Eqs. (\ref{13})-(\ref{9j}), we obtain:
\begin{widetext}
\begin{eqnarray}\label{24}
K(r)=-\frac{\dot{M}}{4\pi r^4}\frac{r}{\sqrt{2U'(r)}} \frac{[2U(r)-r U'(r)][r U''(r)-U'(r)]}{[2U(r)+r U'(r)]^2}
\int^{r}_{mb}Z(r)dr,
\end{eqnarray}
where
\begin{eqnarray}\label{25}
Z(r)=&&\sqrt{\frac{r}{2U'(r)}}\frac{[2U(r)+r U'(r)][-U''(r)r U(r)+2r U'^2(r)-3U'(r)U(r)]}{[2U(r)-r U'(r)]^2}.
\end{eqnarray}
\end{widetext}

In the context of a steady-state accretion disk, which is assumed to be in thermodynamic equilibrium, the radiation emitted by the disk can be characterized as black body radiation. Therefore, we can express the relationship $ K(r) = \sigma T^4(r) $, where $ \sigma $ represents Stefan's constant. A comprehensive analysis of this relationship has been provided by \cite{55f} and subsequently discussed by \cite{53f}.
\begin{figure*}
\centering \epsfig{file=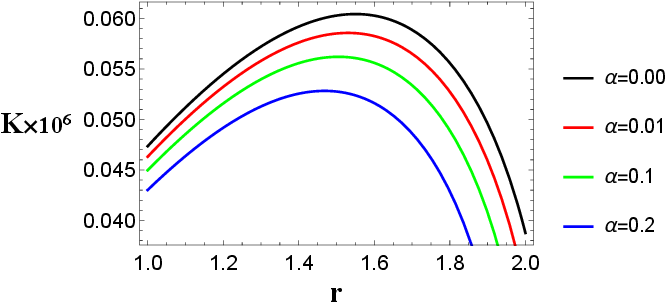, width=.45\linewidth,
height=2.2in}\epsfig{file=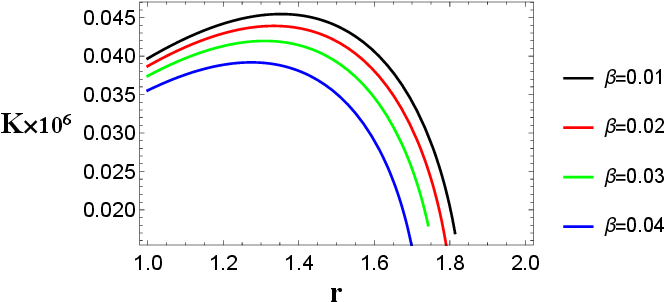, width=.45\linewidth,
height=2.2in}\caption{\label{F20} The radiation flux of decoupled BH.}
\end{figure*}
The radiation flux is depicted in Fig. (\ref{F20}) and has the following key points: In the left plot, the radiation flux decreases and decreases the radius from the killing horizon by the decoupled parameter $\alpha$. In the right plot, the radiation flux decreases and decreases the radius from the killing horizon by the decoupled parameter $\beta$.

The efficiency of the accreting fluid is a crucial aspect of the analysis of the accretion disk. The highest efficiency for converting energy into the radiative flux of particles traveling from the ISCOs to infinity is defined as the ratio of the binding energy at the ISCOs to the rest mass energy. Consequently, the relationships between efficiency and maximum accretion efficiency are expressed as $\chi=1-\eta$ and $\chi^{\ast}=1-\eta_{ISCOs}$, respectively.
\begin{figure*}
\centering \epsfig{file=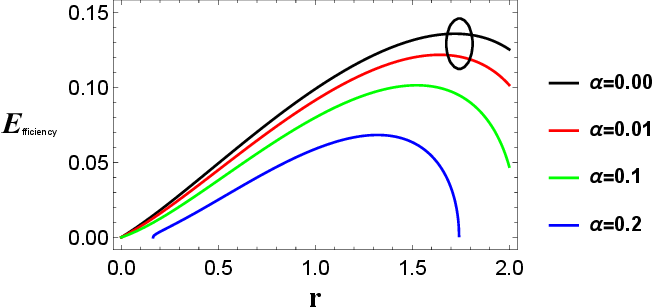, width=.45\linewidth,
height=2.2in}\epsfig{file=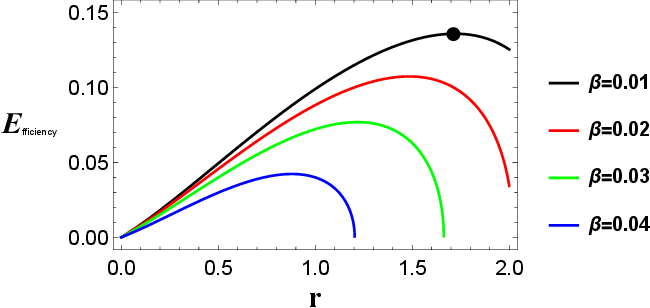, width=.45\linewidth,
height=2.2in}\caption{\label{F21} The energy efficiency of a particle falling from a special infinity into the BH is shown as a function of $ r $ for various values of the decoupled parameters. The peak efficiency of the massive particle at the ISCOs is illustrated by a disk on the black solution curve.}
\end{figure*}
The trend of efficiency is illustrated in Fig. (\ref{F21}), where we observe a decrease as the values of $\alpha$ and $\beta$ increase.
\begin{figure*}
\centering \epsfig{file=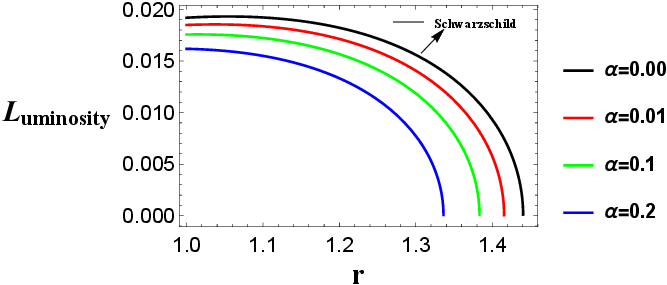, width=.45\linewidth,
height=2.2in}\epsfig{file=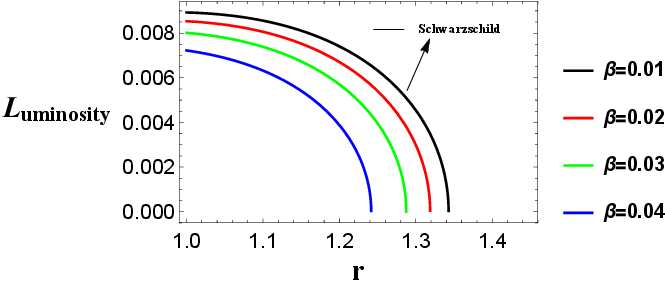, width=.45\linewidth,
height=2.2in}\caption{\label{F22} The observed luminosity at the distance $d$ to the source with
the disk inclination angle $\gamma$.}
\end{figure*}
By utilizing the temperature distribution on the disk and assuming thermal black body radiation, we can calculate the luminosity $L(\nu)$ of the disk. The observed luminosity at a distance $d$ from the source, taking into account the disk's inclination angle $\gamma$, is given by the following expression \cite{50bbff}:
\begin{eqnarray}\label{25b}
L(\nu)=4\pi d^2 I(\nu)=\frac{8}{\pi}(\cos \gamma)\int^{r_f}_{r_i}\int^{2\pi}_{0}\frac{\nu^3_e r d\phi dr}{\exp (\frac{\nu_e}{T})-1}.
\end{eqnarray}
In this context, $I(\nu)$ represents the thermal energy flux. The variable $r_i$ denotes the location of the inner edge, which we set as $r_i = r_{mb}$. Meanwhile, $r_f$ indicates the outer edge of the disk. For any compact general relativistic object, the flux on the disk surface can approach zero as $r \rightarrow \infty$, so we consider $r_f \rightarrow \infty$. The frequency emitted is expressed as $\nu_e = \nu(1+z)$, where:
\begin{eqnarray}\label{25c}
z=\frac{1+\Omega_\varphi r \sin \varphi \sin \gamma}{\sqrt{-U(r)-\Omega^2_\varphi g_{\varphi\varphi}}}-1.
\end{eqnarray}

The luminosity trend is depicted in Fig. (\ref{F22}), where we can see that it decreases as the values of $\alpha$ and $\beta$ increase, and conversely.

\section{Impact of Decoupled Hairy Parameters on Shock Cone Morphology and QPO Structure} \label{GRH1}

The Bondi–Hoyle–Lyttleton (BHL) accretion mechanism is an important tool for testing modified gravity theories, as it allows the investigation of how the parameters of such theories influence observable quasi-periodic oscillation (QPO) signatures and the dynamics of gravity in strong-field regimes \cite{Donmez2014MNRAS,Koyuncu:2014MPLA,Donmez2022Univ,Donmez2024JCAP,Donmez2024Universe,Donmez2024MPLA,DONMEZ2024PDU,Donmez2025JHEAp,Donmez2025EPJC,Mustafa2025JCAP}. In this section, we reveal the effects of the decoupled hairy parameters $\alpha$ and $\beta$ on the shock cone that forms around the BH, thus obtaining potential observational indicators. To achieve this, we solve the general relativistic hydrodynamics (GRHD) equations in conservative form using high-resolution numerical methods, enabling us to capture the morphology of the shock cone that emerges in strong gravitational fields\cite{Donmez2004ASS,Donmez2006AMC,Donmez2017MPLA}. By modeling the fully time-dependent BHL accretion flow as it falls onto the BH, we compute the downstream accretion patterns, the formation of the shock cone, and the oscillation frequencies of the QPO modes trapped within the cone, including their instabilities. The results presented here demonstrate how modifications to the spacetime through $\alpha$ and $\beta$ affect the accretion dynamics and leave measurable imprints on the QPO spectrum, offering a pathway for constraining these parameters observationally.

The general influence of the decoupled hair parameters $\alpha$ and $\beta$ on the  plasma structure and the shock cone formed around the BH is illustrated in Fig. (\ref{color_plots}). The lapse function of the decoupled hair BH, given in  Eq. (\ref{1}), contains the deformation term $\Delta U(r) = \frac{\alpha M}{r^{2}}\left(1+\ln\frac{r}{\beta}\right)$, which is positive and increases with $\alpha$. A larger lapse function compared to the Schwarzschild case at the same radius implies that the spacetime is less redshifted, meaning that the effective potential well becomes shallower. More importantly, the radial derivative of Eq. (\ref{1}) includes the negative correction $-1 - 2\ln(r/\beta) < 0$, which reduces \( U'(r) \) relative to the Schwarzschild value. This decrease in $U'(r)$ weakens the inward gravitational acceleration experienced by infalling matter. Consequently, gravitational focus diminishes, compression in the downstream region is reduced, and the shock cone develops a larger opening angle.

As seen in Fig. (\ref{color_plots}), this behavior is clearly reflected in the rest-mass density contours for each choice of the hair parameters compared to the Schwarzschild geometry. In the Schwarzschild case, the shock cone is narrow and highly compressed, since matter is efficiently focused toward the axis. As  $\alpha$ increases (with $\beta$ fixed), the density contours systematically widen. At an identical downstream location, the high density region spans a greater azimuthal interval, and the corresponding isodensity levels exhibit a noticeably reduced pinching toward the symmetry axis. As further discussed in Fig. (\ref{densty_veloc}), the shock opening angle increases and the color maps show that the density decreases progressively with increasing $\alpha$. This behavior follows directly from the weakening of gravitational focusing. The reduced gravitational pull allows pressure forces to push the shock locations outward, leading to a broader cone.

The velocity vector fields in Fig. (\ref{color_plots}) also support this interpretation. As  $\alpha$ increases, the flow vectors exhibit a noticeably weaker bending toward the BH compared to the Schwarzschild case. The inward directed component of the velocity within the cone is visibly larger for Schwarzschild than for any of the $\alpha>0$ configurations. Fewer streamlines are strongly compressed onto the symmetry axis, while more of them skirt the downstream region at finite polar angle. In other words, the gravitational lensing of the infalling matter becomes weaker for larger $\alpha$, resulting in less focused accretion trajectories, precisely in agreement with the behavior displayed by the density contours.

In general, Fig. (\ref{color_plots}) shows that the opening angle of the shock cone increases with  $\alpha$, the density contours become wider, and the velocity field converges less strongly towards the BH. All of these features are naturally the result of the weakening of the effective gravitational focus induced by the deformation term $\Delta U(r)$.

\begin{figure*}[!ht]
  \vspace{1cm}
  \centering
 \psfig{file=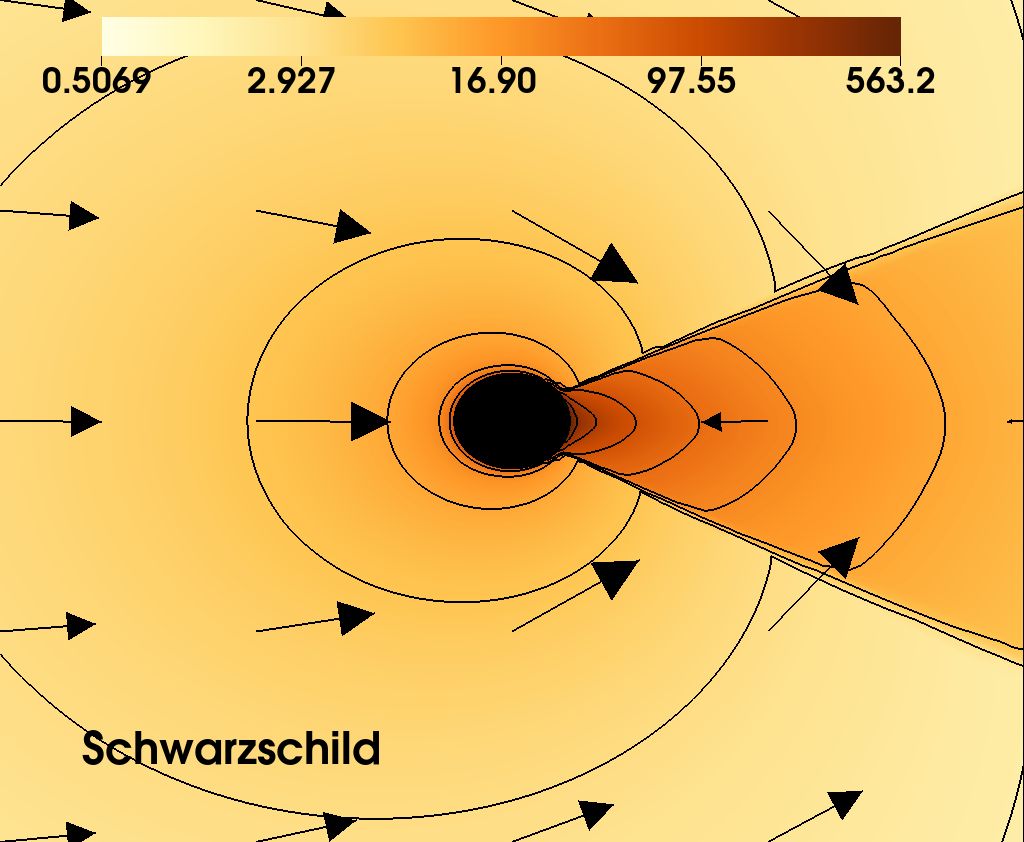,width=7.5cm,height=7cm}
 \psfig{file=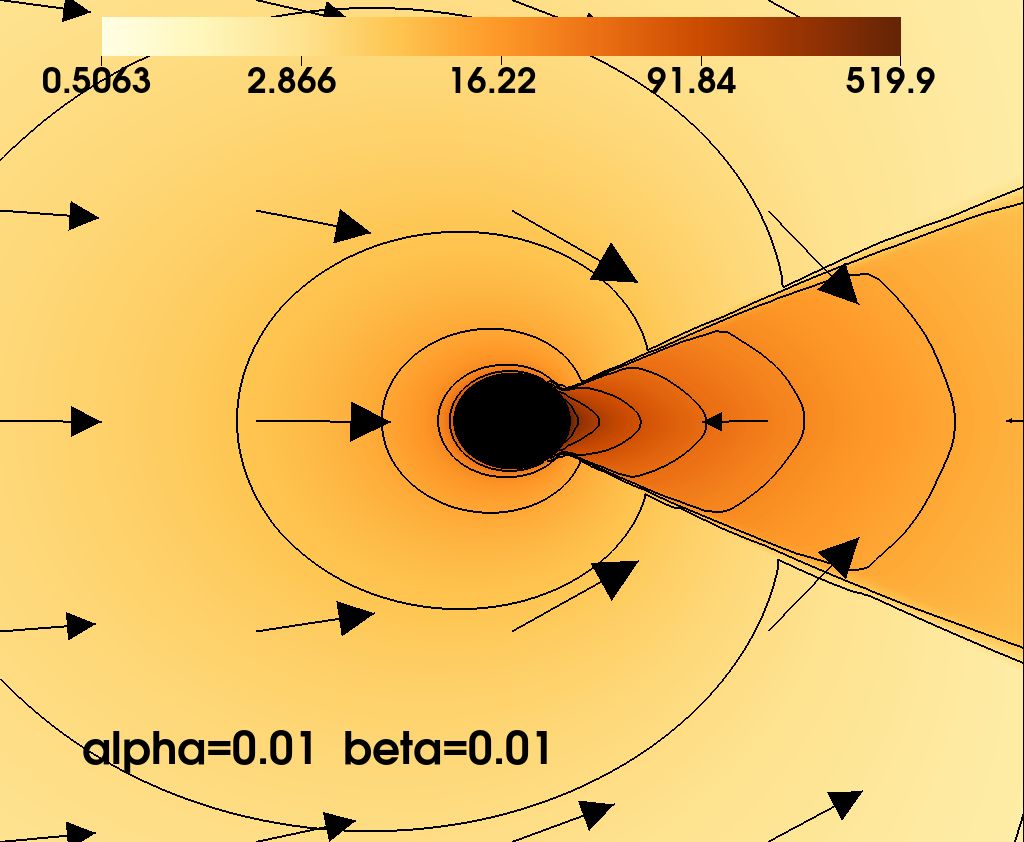,width=7.5cm,height=7cm}\\
 \psfig{file=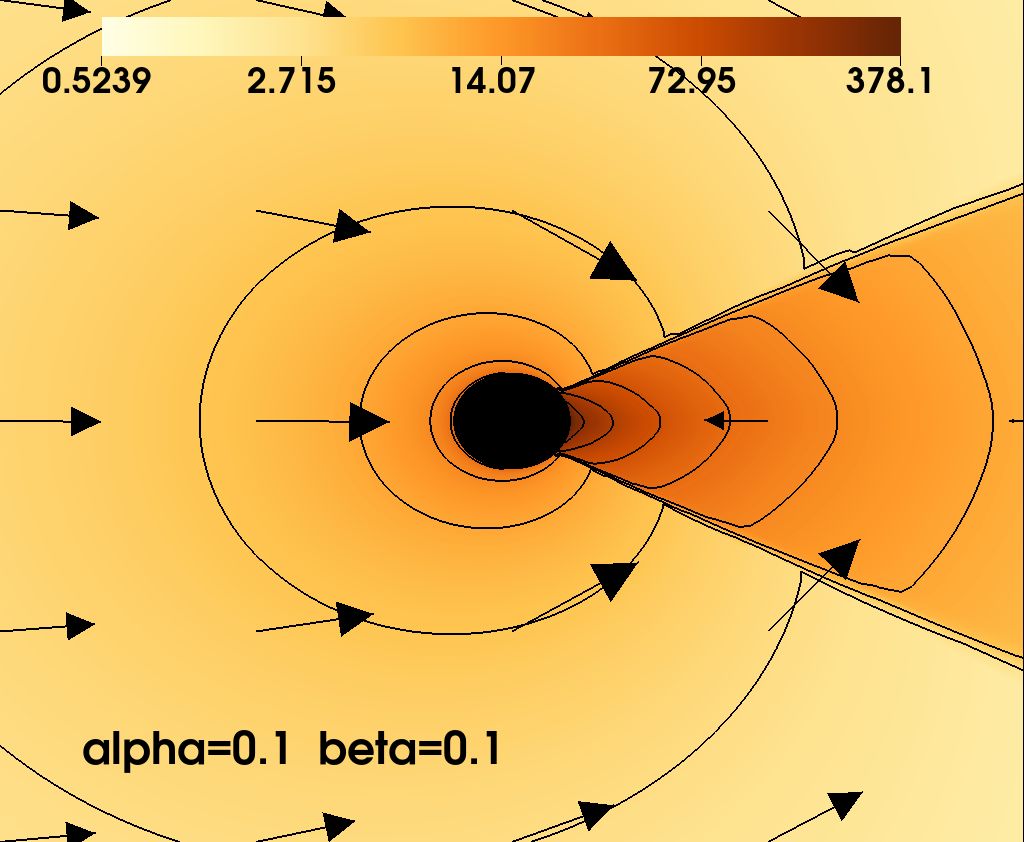,width=7.5cm,height=7cm}
 \psfig{file=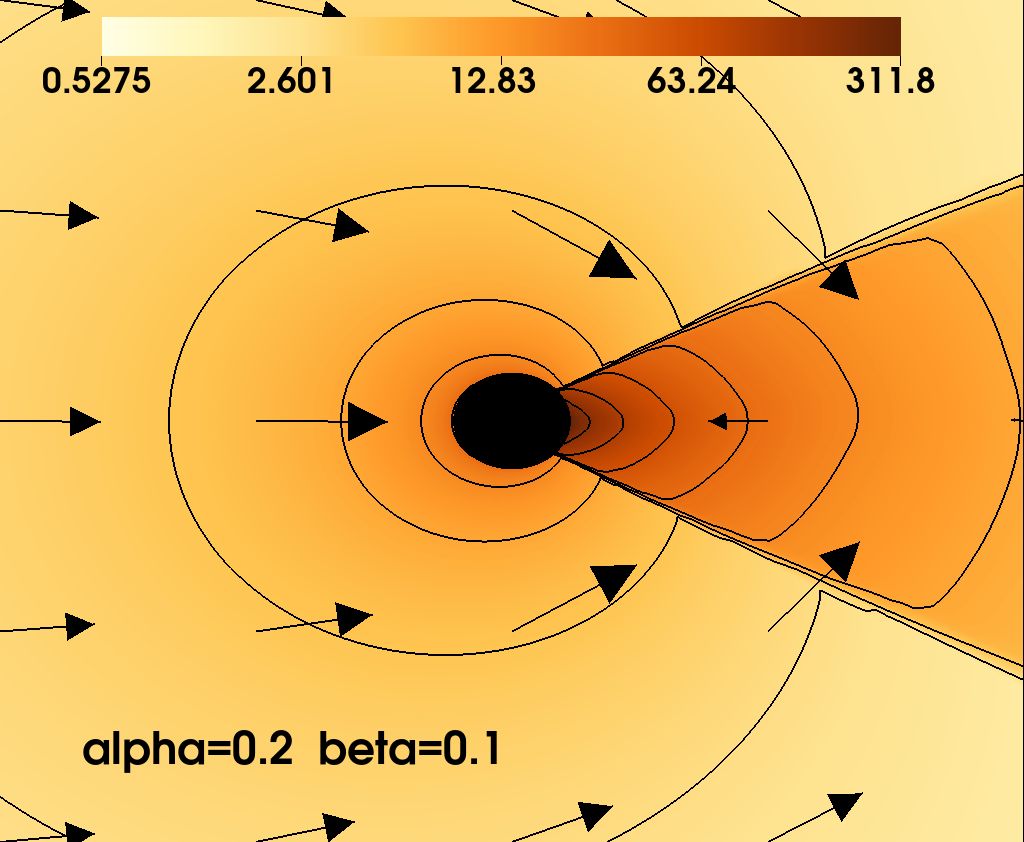,width=7.5cm,height=7cm}
  \caption{The variation of the rest-mass density in the accretion dynamics around the decoupled hairy BH is shown for different values of the hair parameters $\alpha$ and $\beta$, together with the Schwarzschild case. The density distribution is represented using color maps and contour plots, revealing how the contour patterns change from model to model. In addition, the vector plots illustrate the behavior of the infalling matter, demonstrating how the flow slows down within the formed shock cone. To clearly observe the variations occurring near the BH, the computational domain is zoomed to $[x,y] = [20M, 20M]$. Different panels correspond to distinct values of the hair parameters $(\alpha, \beta)$, showing the influence of decoupling on the accretion dynamics.
}
  \vspace{1cm}
  \label{color_plots}
\end{figure*}

Fig.\ref{mass_acc} presents a comparative analysis of the time evolution of mass  accretion rates for Schwarzschild and decoupled hairy BHs at two  diagnostic radial locations,  $r = 2.3M$ (left panel) and $r = 6.11M$  (right panel). In the region very close to the horizon, where the gravitational  field is extremely strong ($r = 2.3M$), all models exhibit pronounced temporal oscillations, which originate from the unsteady dynamics of the shock  cone. As clearly seen in Fig. (\ref{mass_acc}), the average mass accretion rate associated  with the hairy geometries is higher than that of the Schwarzschild case, and the  amplitude of the oscillations becomes increasingly pronounced as the parameter  $\alpha$ increases. This behavior follows directly from the earlier discussion of the lapse function and its derivative. The weakening of the gravitational focus in the strong field region broadens the shock cone opening angle. Although this broadening reduces the axial density peak, it increases the fraction of material that intersects the spherical surface at $r = 2.3M$, which improves the instantaneous flux and strengthens the modulation of the accretion rate.

In contrast, the behavior at $r = 6.11M$, shown in the right panel of Fig.\ref{mass_acc}, is qualitatively different. Near the ISCOs, the support provided by the angular momentum and the detailed structure of the effective potential play an important role in determining the local mass flux. For larger $\alpha$, the reduced gravitational focus leads to less efficient channeling of matter toward the BH, resulting in a lower density and a diminished inflow velocity at this radius. Consequently, the accretion rate in $r = 6.11M$ exhibits a decreasing trend with increasing $\alpha$, and all hairy BH models produce smaller accretion rates than in the Schwarzschild case.

Despite the differences between the two radii, both locations display persistent quasi-periodic behavior in the mass accretion rates. These oscillations indicate that the shock cone and the cavity it forms excite and trap fundamental QPO modes in both the radial and azimuthal directions. Thus, variations in $\alpha$ modify the geometry and dynamics of the shock cone and consequently alter the spectral characteristics of the accretion flow relative to Schwarzschild. As a result, the associated QPO frequencies shift in a predictable way, offering a potential observational avenue for testing the decoupled hairy BH model.

\begin{figure*}[!ht]
  \vspace{1cm}
  \centering
 \psfig{file=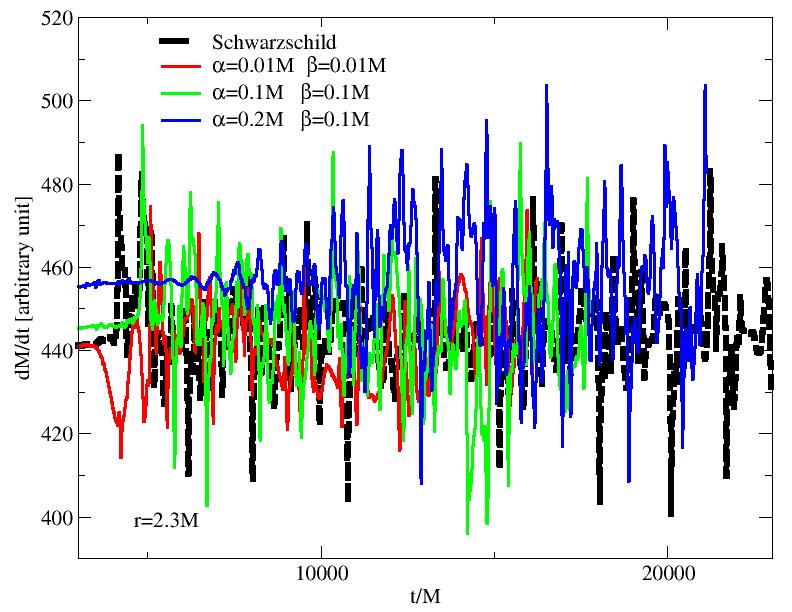,width=7.5cm,height=8cm}
 \psfig{file=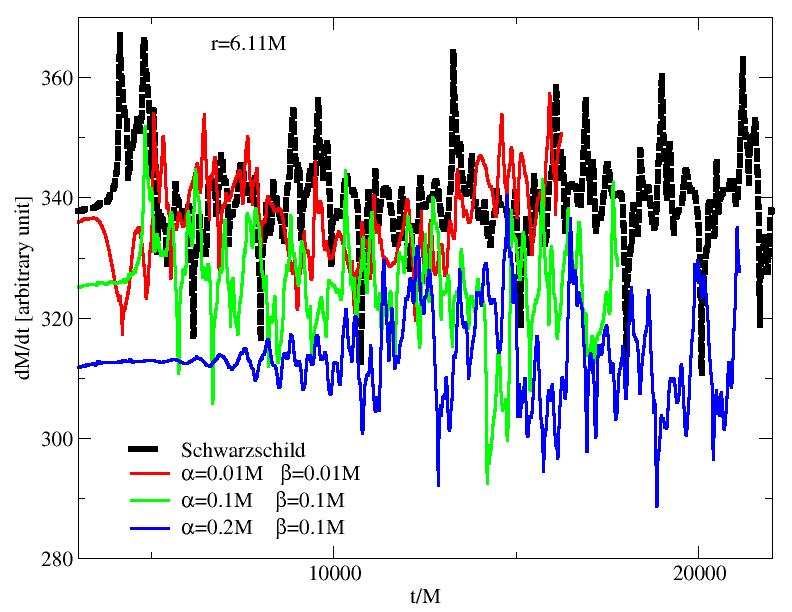,width=7.5cm,height=8cm}\\
  \caption{Comparison of the time evolution of the mass accretion rate for different values of the decoupled hair parameters $(\alpha, \beta)$ with the Schwarzschild case.  The left panel shows the variation of the mass accretion rate at $r = 2.3M$, which is very close to the event horizon, while the right panel presents its behavior at $r = 6.11M$, corresponding to the region near the ISCOs. In both cases, the accretion rate exhibits strong oscillations over time. However, the amplitude of these oscillations becomes more pronounced for higher values of $\alpha$. At $r = 2.3M$, the average accretion rate slightly increases compared to the Schwarzschild case, whereas at $r = 6.11M$ the average rate decreases as $\alpha$ increases.
}
  \vspace{1cm}
  \label{mass_acc}
\end{figure*}

In Fig. (\ref{densty_veloc}), a more quantitative view of the shock cone structure seen in Fig. (\ref{color_plots}) is presented by separating the azimuthal density distribution from the radial kinematics of the matter trapped inside the cone. The left panel shows the azimuthal variation of the density at $r = 2.66M$, very close to the BH horizon, illustrating the cone opening angle and the  redistribution of matter around the downstream axis. As seen in Fig., the  Schwarzschild solution exhibits the narrowest and highest density peak,  indicating that in the Schwarzschild geometry the accreted matter is strongly  compressed, well collimated, and confined to a small azimuthal interval around  the downstream region. The same panel also shows that as $\alpha$ increases, the maximum density of the material trapped inside the cone systematically decreases, while the profile simultaneously broadens in $\phi$. Thus, the cone opening angle increases and the axial overdensity becomes diluted. This behavior is entirely consistent with the two-dimensional density maps of Fig. (\ref{color_plots}). Because the deformation term in the lapse function weakens the effective potential, the fluid trajectories remain less tightly bound to the BH, producing a wider and less compressed shock cone relative to the Schwarzschild case.

The right panel of Fig. (\ref{densty_veloc}) displays the radial velocity $v^{r}/c$ of the matter inside the cone as a function of radius along $\phi = 0$. All models  show a characteristic transition from a nearly static or mildly inflowing gas at large radii to a rapidly plunging flow close to the horizon. The point where $v^{r}/c = 0$ identifies the stagnation point. The matter inside this radius flows toward the BH, whereas the matter outside this radius is advected toward the outer numerical boundary. In the decoupled hairy BH models, the stagnation point is slightly inward, closer to the BH. This implies that the region of net inflow extends deeper into the potential well, thereby modifying both the amount of matter available for accretion and the effective size of the resonant cavity formed between the shock cone and the stagnation surface. Combined with the widening of the shock cone opening angle seen in Figs. (\ref{color_plots} and \ref{densty_veloc}) (left panel), this naturally explains the behavior of the  mass accretion rates shown in Fig. (\ref{mass_acc}), near the horizon at $r = 2.3M$, a wider cone together with an inward shifted stagnation point enhances the inward  flux and increases the oscillation amplitude of $dM/dt$ as $\alpha$ grows. In contrast, in the more distant radius $r = 6.11M$, weakened focusing reduces both the local density and the inflow speed, resulting in a lower average accretion rate for larger $\alpha$.

From a dynamical perspective, the radial velocity profiles in Fig. (\ref{densty_veloc}) also provide information about the transition from quasi-circular motion to plunging orbits, i.e., the region associated with the ISCOs in the underlying geodesic structure. The radius at which $v^{r}$ steepens sharply toward more negative values marks the onset of a rapid fall. In the hairy BH models this turning behavior is slightly shifted relative to Schwarzschild, indicating a displacement of the effective ISCOs region. The QPO modes excited inside the shock cavity depend on both the radial extent of the inflow region and the local dynamical timescales, set by the radial and azimuthal epicyclic frequencies. This geometric and kinematic deformation inevitably modifies the resulting QPO spectrum.

\begin{figure*}[!ht]
  \vspace{1cm}
  \centering
 \psfig{file=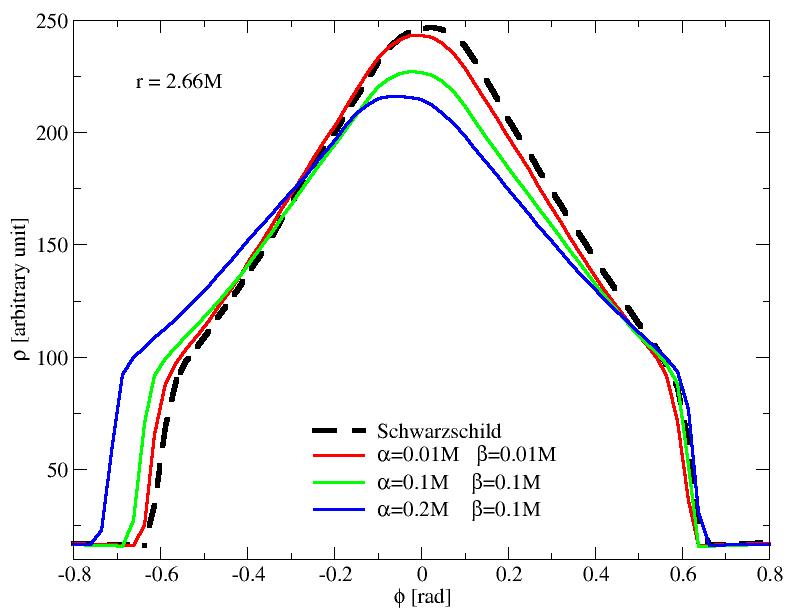,width=7.5cm,height=8cm}
 \psfig{file=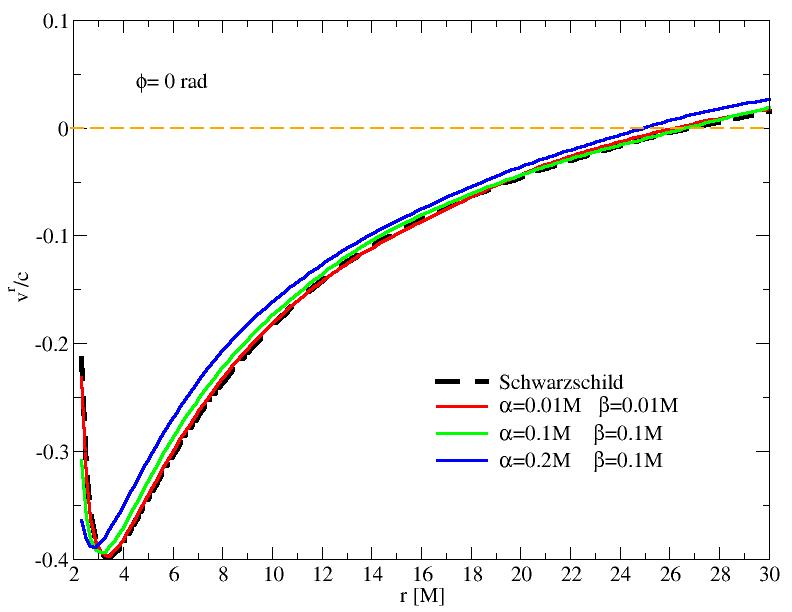,width=7.5cm,height=8cm}\\
  \caption{The structure of the shock cone formed around the decoupled hairy BH and the flow properties of the matter trapped inside the cone are illustrated for different values of the hair parameters $(\alpha, \beta)$, along with the Schwarzschild BH for comparison. The left panel shows the variation of the rest-mass density along the azimuthal direction very close to the black-hole horizon at $r = 2.66M$, comparing different hair parameter configurations with the Schwarzschild case. The right panel presents the variation of the radial velocity of the matter trapped inside the cone as a function of $r$ along the central axis of the cone at $\phi = 0$. In both cases, the physical quantities significantly deviate from the Schwarzschild solutions due to the dependence of the accretion flow on the hair parameters, demonstrating that the presence of the hair modifies the internal structure and dynamics of the shock cone.
}
  \vspace{1cm}
  \label{densty_veloc}
\end{figure*}

Fig. (\ref{QPOs}) shows the PSD analysis obtained from the mass accretion rate oscillations of three different decoupled hairy BH configurations, compared to the Schwarzschild BH. In each panel, the PSD of the Schwarzschild solution is compared to a single $(\alpha,\beta)$ configuration, allowing us to isolate the effect of the hair parameters on the oscillatory behavior of the shock cone. In every snapshot, the QPO peaks are identified as the highest amplitude PSD features, and their frequencies clearly differ from one model to another. This shows that the modified gravity parameters significantly alter the modes trapped inside the shock cavity.

In the left panel of Fig. (\ref{QPOs}) , corresponding to the case $\alpha = 0.01$, $\beta = 0.01$, the dominant peak of QPO shows a slightly reduced amplitude compared to Schwarzschild, but the main peak shifts from the Schwarzschild LFQPO value near $28.5\,\mathrm{Hz}$ to HFQPO at approximately $64.7\,\mathrm{Hz}$. Observational and theoretical studies indicate that LFQPOs generally occur in the $0.1 $-$ 30\,\mathrm{Hz}$ range, while frequencies above $30\,\mathrm{Hz}$ fall into the HFQPO regime \citep{Remillard2006ARAA,Lewin2006Book}. Thus, although the peak amplitude decreases, the QPO frequency nearly doubles in this case. For the intermediate configuration $\alpha = 0.1$, $\beta = 0.1$ (middle panel), the QPO amplitude increases significantly compared to the Schwarzschild BH. However, unlike the previous case, the dominant peak shifts downward, instead of the Schwarzschild peak near $28.5\,\mathrm{Hz}$, the hairy BH produces a main peak at $20.8\,\mathrm{Hz}$, fully consistent with the LFQPO range. In the strongest deformation case, $\alpha = 0.2$, $\beta = 0.1$ (right panel), the QPO frequency undergoes a large shift,  the dominant peak moves from $ 28.5\,\mathrm{Hz} $ down to roughly $9\,\mathrm{Hz}$, squarely within the LFQPO band. Here, the amplitude is also strongly modified, indicating significant restructuring of the trapped modes. The most dramatic percentage change occurs in this model, where the frequency decreases by more than a factor of three relative to Schwarzschild, producing a clearly distinguishable observational signature.

These numerical results demonstrate that the QPO frequencies produced by the shock cavity lie in the $10$-$70\,\mathrm{Hz}$ band, which matches the observed ranges of LFQPO and HFQPO in microquasars. The amplitudes of the dominant modes are also substantially enhanced, suggesting that such signals would be observable with current X-ray timing instruments. The model dependent shifts in QPO frequencies are consistent with the variety of QPO detections from the same astrophysical sources, potentially explaining why different observations of a single system exhibit different QPO frequencies. Overall, the PSD analysis confirms that QPO signatures are sensitive probes of the underlying spacetime geometry, the frequency reductions and amplitude changes induced by the decoupled hairy parameters should be detectable in high quality X-ray observations, offering a realistic pathway for constraining or testing this modified gravity scenario.

\begin{figure*}[!ht]
  \vspace{1cm}
  \centering
 \psfig{file=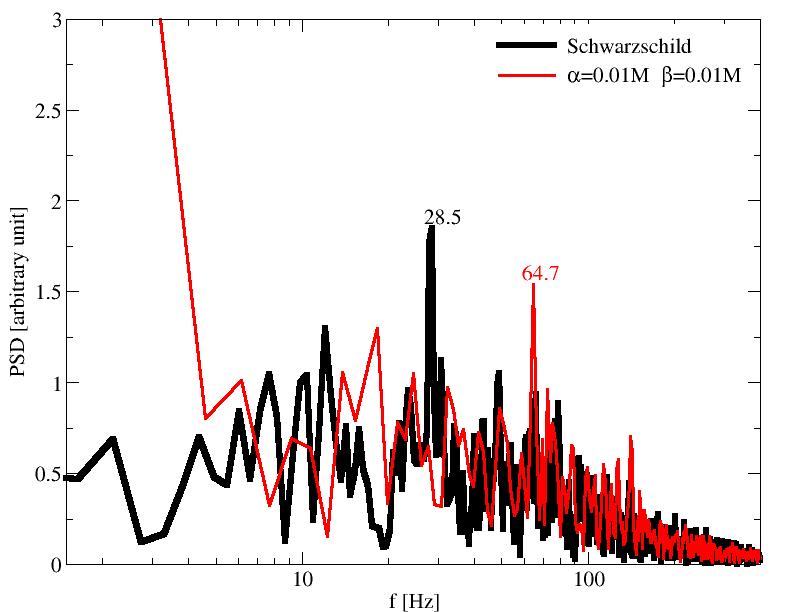,width=5.0cm,height=8cm}
 \psfig{file=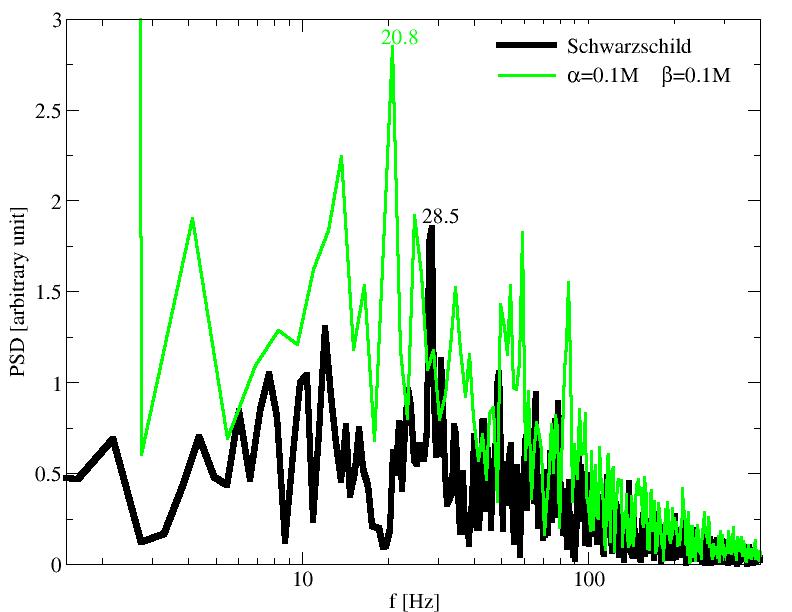,width=5.0cm,height=8cm}
 \psfig{file=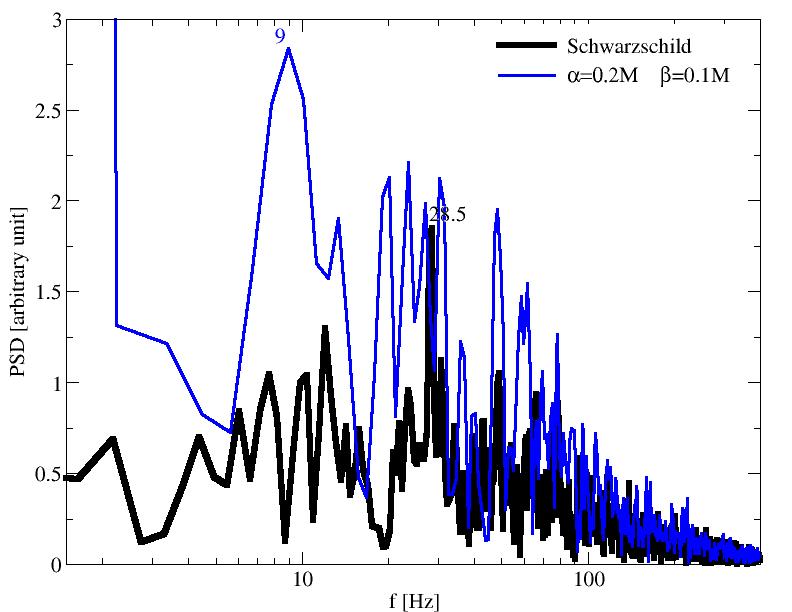,width=5.0cm,height=8cm}\\ 
  \caption{PSD analyses are performed to reveal the oscillation properties of the shock cone formed around the decoupled hairy BHs and the Schwarzschild BH. Each panel presents a direct comparison of the dominant oscillation frequencies obtained from the shock cone around the Schwarzschild BH and those produced for different hair parameter configurations. The results show that the highest-amplitude PSD peaks shift depending on the hair parameters $\alpha$ and $\beta$. This variation indicates that the presence of hair alters the QPO frequencies, which generally tend to decrease as $\alpha$ increases.
}
  \vspace{1cm}
  \label{QPOs}
\end{figure*}

\section{Agreement Between Analytic Predictions and Numerical Simulations}
\label{Anly-Num}

The numerical results obtained in this work are in excellent agreement with the analytic results presented in Sections (\ref{sec2}-\ref{se5}). For the decoupled hairy BH, the analytically modified lapse function $U(r)$ shows that increasing the parameters $\alpha$ and $\beta$ weakens the gravitational potential. This leads to larger circular orbit radii and pushes the ISCOs outward. Consequently, only higher values of the specific angular momentum and specific energy satisfy the conditions required for bound motion. These analytical predictions are fully consistent with the numerically obtained shock cone morphology, including the widening of the shock opening angle, the reduced accumulation of matter inside the cone, and the outward displacement of the entire shock structure as $\alpha$ or $\beta$ increases, behaviors that are naturally expected when gravity becomes weaker. Similarly, the analytic accretion profiles show that the radial inflow velocity decreases, the sonic point shifts outward, and the mass accretion rate decreases as the hairy parameters grow. The GRHD simulations confirm all of these behaviors. After the cone forms, matter trapped inside it falls toward the BH more slowly, and the density maximum near the BH becomes weaker. Furthermore, the analytical prediction that the ISCOs shift to larger radii and that the effective potential minimum becomes shallower is reflected in the numerical QPO analysis, where the dominant frequencies decrease as the hairy parameters increase, demonstrating that the oscillation region moves to larger radii for higher values of $\alpha$. In general, the hydrodynamical simulations strongly reinforce the analytic predictions. Both analytic and numerical results clearly show that increasing the decoupled hairy parameters weakens the gravitational field, pushes characteristic radii outward, suppresses accretion strength, and significantly enhances the likelihood of LFQPO formation, thereby presenting a coherent and mutually supportive physical picture of the system.

\section{Conclusions} \label{se51}

The study tested a novel BH solution derived using the gravitational decoupling method, where the total energy-momentum tensor is divided into a known seed source and an additional matter distribution. In addition, the resulting metric generalizes the Schwarzschild solution by introducing two deformation parameters, $\alpha$ and $\beta$, which modify the lapse function through logarithmic and inverse-square terms. The exact solution reveals that the geometry lacks both inner and outer horizons for any value of the mass $M$, and the parameters $\alpha$ and $\beta$, indicating a distinctive horizon structure unlike conventional BHs. The solution also satisfies the weak energy condition under specific constraints on the deformation function $H(r)$, ensuring its physical viability. The graphic analyzes in Fig. (\ref{F1}) show how variations in $\alpha$ and $\beta$ affect the event horizon and the behavior, while Fig. (\ref{F3}) shows that the magnetic field strength increases with larger $\alpha$ and smaller $\beta$, particularly away from the singularity. These results show the influence of gravitational decoupling in generating viable, hairy BH configurations and physical structures. 

The analysis shows the dynamics of magnetized particles,neutral particles possessing a magnetic dipole moment,moving around a decoupled geometric BH immersed in an external, weak, and uniform magnetic field. By applying the Wald method, the study derives the electromagnetic potential and corresponding field tensor components, and we can show and illustrate the magnetic field as perceived by a co-moving observer. In addition, the interaction between the magnetic moment of the particle and the external field is rigorously captured through the polarization tensor $S^{\mu\nu}$, culminating in a modified Hamilton-Jacobi equation that governs the particle's motion. A key quantity, the magnetic coupling parameter $\eta$, encapsulates the strength of this interaction and plays a central role in determining the effective potential and conditions for stable circular orbits. The results demonstrate that $\eta$ increases with the decoupling parameters $\alpha$ and $\beta$, and that there exists a critical value $\eta_{\text{ext}}$ beyond which stable circular orbits cannot form. The exact equations are derived for the minimum specific energy $\zeta_{\min}$, the angular momentum $L_{\min}$, and the extreme value $\eta_{\text{ext}}$, showing how these quantities vary with the BH parameters. In particular, all curves of $\eta$ converge near $r = 4.3$, suggesting a universal behavior in the coupling strength at that radius. In this case, we can show and illustrate the influence of these results and imply that the magnetic field and underlying geometric modifications influence the motion and stability of neutral magnetized particles near BHs.

The study of circular motions of magnetized particles around a Weyl geometric BH reveals how the decoupling parameters $\alpha$ and $\beta$ influence the orbital dynamics and stability conditions. In addition, the effective potential, which governs the radial behavior of the test particles, increases as $\alpha$ decreases and $\beta$ increases, indicating that the potential barrier becomes steeper and the orbits of the particles shift closer to the singularity. The angular momentum $L$ and the specific energy $\zeta$ are both sensitive to these parameters: $L$ increases with higher values of $\alpha$ and $\beta$, moving the circular orbits outward, while $\zeta$ increases with increasing $\alpha$ and decreasing $\beta$, suggesting stronger binding energy at closer radii. In this case, the effective force acting on the particle becomes stronger with increasing $\alpha$ and $\beta$, implying a more pronounced influence of spacetime curvature and magnetic coupling on the particle's motion. The ISCOs also show the dependence on these parameters, moving closer to the singularity for higher $\alpha$ and lower $\beta$, as explained by the numerical evaluation of the stability condition. Furthermore, particle trajectories illustrate that increasing the geometric or magnetic parameters $c_1$ and $c_2$ leads to wider and more energetic bound orbits, reinforcing the role of the underlying geometry in shaping the motion of magnetized particles. 

The results show the accretion dynamics of various fluids onto Weyl geometric BHs, incorporating both relativistic and thermodynamic effects. By modeling accreting matter as a perfect fluid and employing an isothermal equation of state $p = \omega \rho$, explicit expressions were obtained for radial velocity and energy density profiles, showing that these quantities are highly sensitive to the state parameter $k$, as well as to the decoupling parameters $\alpha$ and $\beta$. The study reveals that for phantom-like fluids ($k < -1$), the accretion flow exhibits negative radial velocity, implying an outward movement or repulsion near the BH, whereas for ordinary matter ($k \geq 0$), the fluid is accreted inward with increasing velocity. The energy density behaves similarly, decreasing for phantom fields and increasing for conventional fluids as the fluid approaches the BH. In this case, at the sonic point, where the fluid velocity matches the local sound speed, critical conditions were derived, providing insight into the transonic nature of the flow and its stability. The accretion rate and time-dependent BH mass were also calculated, indicating that the BH gains mass from ordinary matter accretion but may lose mass in the presence of phantom energy. In addition, the study tested electromagnetic emissivity and radiation flux from thin accretion disks, showing that both radiation and luminosity decline with increasing values of $\alpha$ and $\beta$, suggesting that the Weyl geometric effects can significantly suppress observable signatures of accretion.

The Figs. in this study collectively illustrate how the dynamical, magnetized and accreting behaviors of particles near a decoupled hairy BH are influenced by the parameters $\alpha$, $\beta$, $M$, $L$, $\zeta$, $\eta$, $k$, $r$, $c_1$, and $c_2$. Fig. (\ref{F1}) displays the variation of the lapse function $U(r)$ with respect to three parameters: the mass of the BH $M$, the decoupling parameter $\alpha$ and the hairy parameter $\beta$. The top-left plot shows curves for $M = 0.1, 0.38, 1, 2$ with fixed $\alpha = 0.001$, $\beta = 0.01$, the top-right plot shows $\alpha = 0.000, 0.001, 0.01, 0.02$ with fixed $M = 1$, $\beta = 0.01$, and the bottom plot shows $\beta = 0.01, 0.02, 0.03, 0.04$ with fixed $M = 1$, $\alpha = 0.001$. In all cases, higher values of the parameters $M$ and $\alpha$ deepen the gravitational potential, while changes in $\beta$ influence the distortion of the horizon structure. Fig. (\ref{F3}) indicates the magnetic field component $G_{\hat{\theta}}$ versus the radial coordinate $r$ for different values of $\alpha = 0.000, 0.01, 0.1, 1$ (left plot) and $\beta = 0.01, 0.02, 0.03, 0.04$ (right plot), showing that $G_{\hat{\theta}}$ increases as $\alpha$ increases and decreases as $\beta$ increases, indicating that the decoupling parameters influence the magnetic topology of the surrounding geometry. Fig. (\ref{F4}) shows how the magnetic coupling parameter $\eta$ varies with radius $r$, for different values of $\alpha = 0.001, 0.01, 0.1, 0.15$ (left plot) and $\beta = 0.01, 0.1, 0.2, 0.3$ (right plot), with fixed $L = 5$ and $\zeta = 1$. The parameter $\eta$ increases with both $\alpha$ and $\beta$, and all curves converge at $r \approx 4.3$. Fig. (\ref{F5}) illustrates the minimum specific energy $\zeta_{\text{min}}$ of the magnetized particles with respect to $r$, with variations in $\alpha = 0.00, 0.01, 0.05, 0.09$ (left plot) and $\beta = 0.01, 0.1, 0.2, 0.3$ (right plot). As $\alpha$ increases and $\beta$ decreases, the value of $\zeta_{\text{min}}$ moves closer to the singularity. Fig. (\ref{F6}) shows the minimum magnetic coupling $\eta_{\text{min}}$ with varying $\alpha = 0.001, 0.01, 0.1, 0.2$ (left plot) and $\beta = 0.01, 0.02, 0.03, 0.04$ (right plot), indicating that $\eta_{\text{min}}$ increases with both $\alpha$ and $\beta$ and shifts away from the singularity. Fig. (\ref{F7}) shows the minimum angular momentum $L_{\text{min}}$ as a function of $r$, for $\alpha = 0.001, 0.01, 0.1, 0.2$ (left) and $\beta = 0.01, 0.02, 0.03, 0.04$ (right). As $\alpha$ decreases and $\beta$ increases, $L_{\text{min}}$ increases and moves farther away from the singularity. Fig. (\ref{F8}) presents the extreme magnetic coupling parameter $\eta_{\text{ext}}$ versus $r$ for $\alpha = 0.001, 0.01, 0.1, 0.2$ and $\beta = 0.01, 0.02, 0.03, 0.04$. In both cases, $\eta_{\text{ext}}$ increases with parameters and asymptotically approaches 1 as $r \rightarrow \infty$, setting a limit for stable circular orbits.

In this case, Fig. (\ref{F9}) plots the effective potential $V_{\text{eff}}$ of magnetized particles as a function of $r$ for values of $\alpha = 0.001, 0.01, 0.1, 0.2$ and $\beta = 0.01, 0.02, 0.03, 0.04$. As $\alpha$ decreases and $\beta$ increases, $V_{\text{eff}}$ increases and becomes more localized near the singularity. Fig. (\ref{F10}) illustrates the angular momentum $L$ as a function of $r$ for $\alpha = 0.001, 0.01, 0.1, 0.2$ (left) and $\beta = 0.01, 0.02, 0.03, 0.04$ (right), showing that $L$ increases with both parameters. Fig. (\ref{F11}) shows the specific energy $\zeta$ against $r$, again for $\alpha = 0.001, 0.01, 0.1, 0.2$ and $\beta = 0.01, 0.02, 0.03, 0.04$. Here, $\zeta$ increases with larger $\alpha$ and smaller $\beta$. Fig. (\ref{F12}) plots the effective radial force $F_{\text{eff}}$, using the same sets of $\alpha$ and $\beta$, demonstrating that the attractive force grows with both parameters. Fig. (\ref{F13}) visualizes the radius of the ISCOs $r_{\text{ISCOs}}$ for fixed $\eta = 0.01$, with respect to $\beta = 2.5, 3.0, 3.5, 4.0, 4.5, 5.0$ (left) and $\alpha = 0.001, 0.01, 0.1, 0.2$ (right). The ISCOs radius parameters decrease (moves closer to the BH) with increasing $\alpha$ and decreasing $\beta$. Figs. (\ref{F14}) and (\ref{F15}) depict the trajectories of spinning magnetized particles under the influence of Weyl parameters $c_1$ and $c_2$. Increasing these parameters expands the orbital radius and energy of the particle, indicating that the geometry's curvature strongly influences the orbital structure. In addition, in the final section on accretion, Fig. (\ref{F16}) shows the radial velocity $u(r)$ for various equations of state parameterized by $k = -1.5, -1, -0.5, 0, 0.5, 1$. Fluids with $k < 0$ (phantom, quintessence) show a negative radial velocity (infall), while fluids with $k \geq 0$ (dust, radiation, stiff) exhibit a positive velocity. Fig. (\ref{F17}) presents the energy density $\rho(r)$ for the same values of $k$, showing positive density profiles for $k \geq 0$ and negative profiles (non-physical) for phantom-like fluids. Finally, Fig. (\ref{F18}) shows the rate of change in the mass of BH $\dot{M}_{\text{acc}}$ with respect to $r$, for $k = -2, -1.5, -0.5, 0, 0.5, 1, 1.5$. Mass accretion increases with increasing $k$, indicating that stiff and dust fluids contribute more effectively to the growth of BH, while phantom fluids cause a reduction or even a decrease in mass. Across all Figs., the role of the parameters $\alpha$, $\beta$, $M$, $L$, $\zeta$, $\eta$, $k$, $c_1$, $c_2$ and $r$ is clearly shown to significantly control the geometric, dynamical, and energetic behavior of both particles and fluids in this hairy, decoupled spacetime of BH.

In Fig. (\ref{F19}), the emissivity is plotted as a function of the radial coordinate $r$ for fixed $\beta = 0.01$, with varying $\alpha = 0.001, 0.01, 0.1, 0.2$. The increase in $\alpha$ is observed to enhance the emissivity peak and shift it outward. Fig. (\ref{F20}) shows the emissivity profile for fixed $\alpha = 0.01$ and varying $\beta = 0.01, 0.02, 0.03, 0.04$. In this case, higher values of $\beta$ lead to suppression of emissivity and a slight inward shift of the peak, reflecting a weaker geometric deformation of the BH. Fig. (\ref{F21}) explores the effect of the fluid state parameter $k = -1.5, -0.5, 0, 0.5, 1$ for fixed $\alpha = 0.01$ and $\beta = 0.01$. In addition, the results show that more negative values of $k$ (representing phantom-like fluids) increase the emissivity amplitude and bring the peak closer to the BH, while higher values (stiff fluid) reduce the emissivity. In this analysis, Fig. (\ref{F22}) tests the case of a phantom fluid with $k = -1.5$, $\beta = 0.01$ and varying $\alpha = 0.001, 0.01, 0.1, 0.2$, showing that an increase in $\alpha$ significantly amplifies emissivity and extends its influence over a larger radial region. 

The numerical results have shown that the parameters $\alpha$ and $\beta$ have a strong influence on both the QPO behavior and the morphology of the shock cone formed around the hairy BH and that this influence is of an observable nature. As the value of $\alpha$ increases, the gravitational potential deepens, causing the opening angle of the shock cone to widen and the cone itself to become more asymmetric. This effect arises entirely from the density gradients generated by the hydrodynamic forces within the accretion mechanism. Compared with the Schwarzschild case, these morphological changes lead to noticeable shifts in the QPO frequencies. The magnitude of these shifts depends directly on the strength of the hairy parameters. Thus, the resulting QPO structures clearly encode signatures of the underlying accretion geometry around the BH. This demonstrates that the hairy parameters $\alpha$ and $\beta$ leave observable imprints on the timing properties of the accreting matter, potentially serving as powerful probes for testing hairy BH geometries in future X-ray and gravitational wave observations.

\section*{Acknowledgments}
All numerical simulations were performed using the Phoenix High
Performance Computing facility at the American University of the Middle East (AUM), Kuwait.

\end{document}